\documentclass[fleqn,usenatbib]{mnras}

\usepackage{mathptmx}
\usepackage{multicol}
\usepackage{subcaption}
\usepackage{color}

\usepackage[T1]{fontenc}

\DeclareRobustCommand{\VAN}[3]{#2}
\let\VANthebibliography\thebibliography
\def\thebibliography{\DeclareRobustCommand{\VAN}[3]{##3}\VANthebibliography}

\usepackage{graphicx}	
\usepackage{amsmath}	
\usepackage{soul}

\newcommand{\teff}{$T_{\rm eff}$}
\newcommand{\vmic}{$\xi_\mathrm{mic}$}
\newcommand{\logg}{$\log g$}
\newcommand{\feh}{[Fe/H]}

\newcommand{\mdot}{$M_\odot$}
\newcommand{\cd}{d$^{-1}$}
\newcommand{\luminosity}{$\log (L/{\rm L_\odot})$}
\newcommand{\kms}{km\,s$^{-1}$}
\newcommand{\vsini}{$\upsilon\sin i$}

\newcommand{\frot}{f$_{\rm rot}$}

\newcommand{\tess}{{\it TESS}}

\newcommand{\logAX}{$\log {\rm (A)}_{X}$}

\definecolor{blu}{rgb}{0,0,1}
\definecolor{mag}{rgb}{1,0,1}
\definecolor{dgreen}{rgb}{0.1, 0.53, 0.22}
\definecolor{orange}{RGB}{255,127,0}
\definecolor{debianred}{rgb}{0.84, 0.04, 0.33}

\title[The Ap Si/He-wk star HD\,100357]{Chemical abundances and doppler imaging of the Ap Si/He-wk star HD\,100357}

\author[A. Dileep et al.]{
Athul Dileep$^{1,2}$,\thanks{E-mail: dileep@aries.res.in}
Santosh Joshi$^{1}$, 
Sofya Alexeeva$^{3}$,
Oleg Kochukhov$^{4}$,
Eugene Semenko$^{5}$,
Peter De Cat$^{6}$,
\newauthor
Sebastián Zúñiga-Fernández$^{7}$,
Otto Trust$^{8, 9}$,
Karen Pollard$^{10}$,
Lisa Crause$^{11}$,
Khalid Barkaoui$^{7,12,13}$,
\newauthor
Michaël Gillon$^{7}$,
Emmanuel Jehin$^{14}$,
and Neeraj Rathore$^{15}$
\\
$^{1}$Aryabhatta Research Institute of Observational Sciences, Manora Peak, Nainital 263002, Uttarakhand, India\\
$^{2}$Department of Applied Physics, M.J.P. Rohilkhand University, Bareilly 243006, Uttar Pradesh, India \\
$^{3}$National Astronomical Observatories, Chinese Academy of Sciences, A20 Datun Road, Chaoyang, Beijing 100101, China\\
$^{4}$Department of Physics and Astronomy, Uppsala University, Box 516, 751 20
Uppsala, Sweden\\
$^{5}$National Astronomical Research Institute of Thailand (NARIT), 260, Moo 4, Don Kaew, Mae Rim, Chiang Mai, Thailand, 50180\\
$^{6}$Royal Observatory of Belgium (ROB), Av. Circulaire 3, 1180 Uccle, Belgium\\
$^{7}$Astrobiology Research Unit, Universit\'e de Li\`ege, All\'ee du 6 Ao\^ut 19C, B-4000 Li\`ege, Belgium  \\
$^{8}$Physics Department, Mbarara University of Science and Technology, P.O. Box 1410, Mbarara, Uganda\\
$^{9}$Department of Physics, Faculty of Science, Kyambogo University, P.O. Box 1, Kyambogo, Uganda\\
$^{10}$University of Canterbury, Christchurch 8041, New Zealand\\
$^{11}$South African Astronomical Observatory, Cape Town, 7925, South Africa\\
$^{12}$Instituto de Astrof\'isica de Canarias (IAC), Calle V\'ia L\'actea s/n, 38200, La Laguna, Tenerife, Spain\\
$^{13}$Department of Earth, Atmospheric and Planetary Science, Massachusetts Institute of Technology, 77 Massachusetts Avenue, Cambridge, MA 02139, USA\\
$^{14}$Space Sciences, Technologies and Astrophysics Research (STAR) Institute, Universit\'e de Li\`ege, All\'ee du 6 Ao\^ut 19C, B-4000 Li\`ege, Belgium \\
$^{15}$Bareilly College, Bareilly 243005, Uttar Pradesh, India 
}

\date{Accepted XXX. Received YYY; in original form ZZZ}

\pubyear{2025}

\begin{document}
\label{firstpage}
\pagerange{\pageref{firstpage}--\pageref{lastpage}}
\maketitle

\begin{abstract}
We present the results of time-resolved photometry, abundance analysis and Doppler imaging of an Ap star, HD\,100357. The \tess\ photometry revealed rotational modulation with a period of 1.6279247 days. Upon inspecting the residuals after removing the rotational period and its harmonics, we found additional frequencies around 15.8054 \cd\ which we later confirmed with ground-based observations as originating from a nearby star. Using high-resolution spectroscopy, we identified HD\,100357 as an Ap Si/He-wk star exhibiting rotational modulation caused by surface abundance spots. The stellar parameters of HD\,100357 were determined as \teff\ = 11,850 K, \logg\ = 4.57, \vsini\ = 60 \kms\, and an inclination angle $i$ = 72$^{\circ}$. The detailed abundance analysis revealed strongly overabundant stratified silicon, an overabundance of iron-peak elements and rare earth elements combined with remarkably deficient helium. Mapping of Fe and Cr abundances revealed the existence of ring-shaped regions with a lower concentration of the elements. Their geometry might reflect the orientation of the hypothetical magnetic field of the star, oriented $\sim$90$^{\circ}$ to the rotational axis. HD\,100357, with its strong chemical peculiarities and indications of possible magnetic fields, represents an interesting candidate for follow-up spectropolarimetric observations aimed at investigating its magnetic field topology and stellar activity.
\end{abstract}




\section{INTRODUCTION}

Approximately 10\% of main-sequence B- and A-type stars exhibit strong, globally organised magnetic fields (up to $\sim$30\,kG) and unusually slow rotation (\vsini\,$< 100$\,\kms) \citep{2019MNRAS.483.3127S}, while some exceptions have faster rotations \citep{2022A&A...668A.159M}. These stars, known as magnetic chemically peculiar (CP) stars, display surface abundance anomalies that correlate with effective temperature. The cooler CP stars, classified as Ap SrCrEu stars (7000\,K $\leq$\,\teff\,$\leq$\,10\,000\,K), show pronounced overabundances of Sr, Cr, and rare earth elements, whereas the hotter Ap Si and He-weak stars (10\,000\,$\leq$\,\teff\,$\leq$\,15\,000\,K) exhibit significant Si enhancements and He deficiencies \citep{land2007, romanyuk2013}.

\citet{preston1974} divided CP stars into four main categories: Am (CP1), Ap (CP2), HgMn (CP3), and He-weak (CP4). Of these, CP2 and CP4 stars possess strong global magnetic fields, with strengths typically ranging from a few hundred gauss to several tens of kilogauss \citep{2007A&A...475.1053A, 2010MNRAS.401L..44E, shultz2019}. In contrast, CP1 and CP3 stars are of non-magnetic nature. He-weak stars, in particular, are typically early B-type stars that exhibit strong underabundances of helium and overabundances of elements such as Si, P, and Ga \citep{norris1971, zboril1997}. The He deficiency in these stars is believed to arise from helium gravitational settling and inhibited mixing in the presence of magnetic fields \citep{vauclair1975}.

The anomalies in CP stars are explained by atomic diffusion, in which radiative levitation of certain heavy ions and gravitational settling of lighter ions occur in the absence of convective mixing \citep{michaud1970, richard2001, aleciann2018, alecian2018}. The quenching of mixing in magnetic stars is thought to result from magnetic braking that slows stellar rotation during early evolutionary stages \citep{1980ApJ...240..218F}, promoting atmospheric stability.

The magnetic fields in CP stars not only govern diffusion but also suppress differential rotation and convective turbulence, producing horizontal abundance patches across the stellar surface. These spots rotate with the star, giving rise to photometric and spectroscopic variability on rotational timescales. Such variability can be used to infer stellar rotation periods and map surface inhomogeneities via Doppler Imaging (DI) and Zeeman Doppler Imaging (ZDI; \citealt{kochukhov2004, donati2006}). Rotation periods of magnetic CP stars span a wide range, from less than a day to several decades, with very long periods associated with the most strongly magnetic examples \citep{mathys2017, 2022MNRAS.514.3485G,  2024A&A...691A.186M}.

In addition to surface inhomogeneities, vertical stratification in the stellar atmosphere causes different chemical elements to concentrate at different depths. This enables the use of lines forming at different atmospheric layers to probe depth-dependent pulsation modes, particularly in a cooler subset of the magnetic CP stars that pulsate in high-overtones, low-degree, non-radial $p$-modes aligned with their magnetic axes, known as the rapidly oscillating Ap (roAp) stars \citep{1998MNRAS.300L..39B, 2001A&A...374..615K, 2003MNRAS.345..781M, ryabchikova2004}. The strong chemical peculiarities and magnetic field in the presence of pulsations provide great insights into stellar interiors \citep{2015JApA...36...33J, 2022afas.confE...1K} and are among the best laboratories for studying magneto-acoustic interactions in stellar atmospheres \citep{kurtz2000, saio2005}.

The Nainital–Cape (N-C) Survey is one of the most extensive ground-based campaigns aimed at detecting and characterizing rapidly oscillating Ap (roAp) stars in the northern and southern hemispheres. Using the 1.04-m Sampurnanand Telescope at ARIES, Nainital, and the 0.5-m telescope at the South African Astronomical Observatory (SAAO), hundreds of targets were monitored over nearly two decades. The key results of this systematic survey are summarised in \citet{2000ashoka, 2001martinez, joshi2003, 2006joshi, joshi2009, joshi2010, joshi2012, joshi2016, joshi2017, joshi2022} and have contributed significantly to the discovery of new roAp candidates and the understanding of the low-amplitude variability in CP stars. We continue this study by investigating the photometric variability in CP star candidates from the N-C survey using space-based missions. 

In this paper, we present a comprehensive investigation of the southern Ap Si/He-wk star HD\,100357 (reported a null result for pulsation in the N-C survey) using both ground-based photometric and spectroscopic observations, as well as space-based data from the \tess\ mission. This work expands the sample of Ap Si/He-wk stars in the southern sky, which serve as a transitional class between CP2 and CP4 stars, providing insights into the interactions among magnetism, rotation, and chemical stratification in CP stars.

The manuscript is structured as follows. Sec.\,\ref{data} discusses the observations and data preparation. In Sec.\,\ref{parameters}, we determine the stellar parameters for the star. The chemical compositions and determination of the surface distribution of Fe and Cr are outlined in Sec.\,\ref{abundance} and Sec.\,\ref{Doppler}, respectively. Finally, the summary and conclusion drawn from our findings are given in Sec.\,\ref{conclusion}.

\section{OBSERVATIONS AND DATA ANALYSIS}\label{data}

\subsection{Photometry} 
\subsubsection{\tess} 
\tess\ is an all-sky survey telescope whose primary objective is to detect exoplanets orbiting nearby bright stars using the transit method \citep{2015JATIS...1a4003R}. In addition to the primary mission, \tess\ also provides high-precision time-series photometry suitable for studying other types of variability, such as rotational, pulsational, or eclipses of stars brighter than $12^{th}$ magnitude \citep{2015ApJ...809...77S, 2022MNRAS.513..526B}. \tess\ Target Pixel Files (TPFs) and Light Curve Files are available in short cadences of 20 and 120\,s, while Full-Frame Images (FFIs) are obtained in long cadences: 1800\,s for Sectors 1-26 (Cycles 1-2), 600\,s for Sectors 27-55 (Cycles 3-4), and 200\,s from Sector 56 onwards (Cycle 5+). These data products are publicly available and can be downloaded through the Mikulski Archive for Space Telescopes (MAST){\footnote{\url{https://archive.stsci.edu/}}}.

HD\,100357 was observed in six different sectors by \tess\, and the log of observations is given in Table\,\ref{tess_log}. The \tess\ light curve files contain two types of flux, namely, the Simple Aperture Photometry (SAP) flux and the Pre-search Data Conditioning SAP (PDCSAP) flux. In the PDCSAP flux, long-term instrumental trends are removed using Co-trending Basis Vectors (CBVs) \citep{2012PASP..124.1000S}. For the present study, we used 120\,s, 600\,s, and 1800\,s cadence PDCSAP flux downloaded from the MAST archive via the \textsc{lightkurve} Python package \citep{2018ascl.soft12013L}. 

\begin{table}
\centering
\begin{minipage}{\linewidth}
\caption{Observational log of TESS data. The cadence and duration of observations are given in seconds (s) and days (d), respectively.}
\label{tess_log}
\end{minipage}
\bigskip
\begin{tabular}{ccccc}
\hline
 \textbf{TESS }  &\textbf{Observation Period} & \textbf{Cadence} & \textbf{Duration} \\
 \textbf{Sector}          &              &    (s)          & (d)            \\
\hline

10  & March 26 – April 21, 2019 & 120 & 25.27\\
11  & April 22 – May 20, 2019 &  1800 & 26.04 \\
37  & April 2 – April 28, 2021 &  600 &  24.15\\
38  & April 29 – May 26, 2021 &  600 & 25.74\\
64  & April 6 – May 4, 2023 &  120 & 26.29\\
65  & May 4 – June 1, 2023 &  120 & 27.25\\

\hline
\end{tabular}
\end{table}

\begin{figure}
    \includegraphics[width=\linewidth]{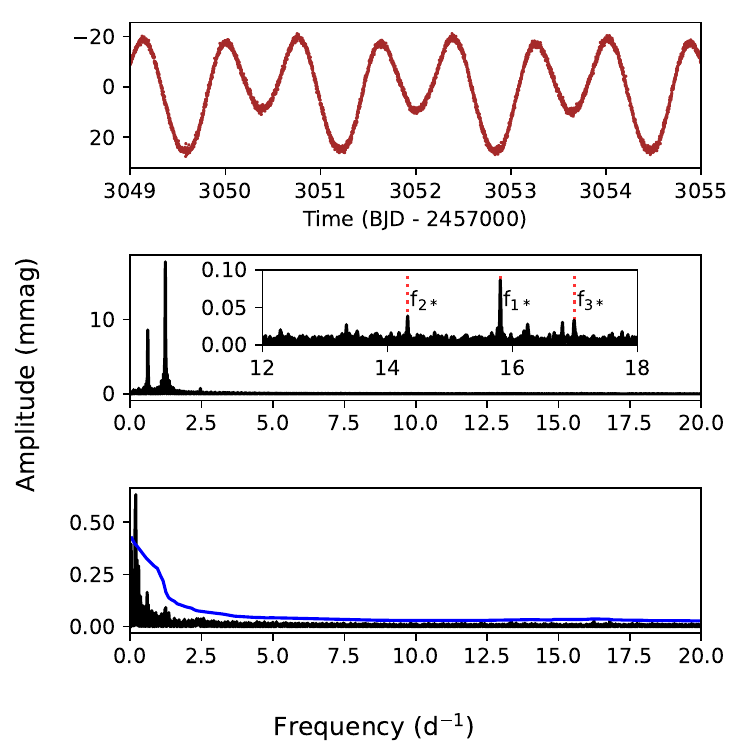}
    \caption{The top panel displays a segment of the TESS light curve with a cadence of 120\,seconds. The second panel shows the amplitude spectrum using the combined light curve from all TESS sectors. The inset in the second panel shows the frequency range from 12 to 18 \cd, where contaminating signals in the residuals persist after fitting and eliminating 15 successive harmonics of the fundamental rotational frequency. The third panel shows the residual after removing all the significant frequencies, where the blue line represents the 5.3\,S/N level. The low-frequency peaks in the residuals can be instrumental artefacts. 
    }
    \label{fig:lc_fourier}
\end{figure}

The light curve fluxes were converted to relative magnitudes (mmag) and centred at zero by subtracting the mean magnitude. Light curves were then stitched together and subjected to frequency analysis to identify the rotational frequency using the discrete fourier fitting technique implemented in the \textsc{Period04} package \citep{2005CoAst.146...53L, 2014ascl.soft07009L}. We fitted the fundamental rotational frequency and its 15 consecutive integer multiples (harmonics) simultaneously using a multi-component sinusoidal model and performed subsequent prewhitening to identify any higher frequencies. A signal-to-noise ratio (S/N) greater than 5.3 was adopted as the criterion for selecting significant frequencies, based on the detection threshold for \tess\, sectors by \citet{2021AcA....71..113B}. The derived frequencies, amplitudes, and S/N values obtained are listed in Table\,\ref{Tab:Table1}. The light curve, frequency spectrum and residuals are shown in Fig.\,\ref{fig:lc_fourier}. 

HD\,100357 (TIC\,280667311) was classified as an Ap star with A0 spectral type and showing EuCrSr type peculiarity by \citet{1975mcts.book.....H}. These stars are known to show rotational variability due to chemical abundance spots. \citet{2020MNRAS.493.3293B} reported a rotational period of 1.62794\,d for HD\,100357 star using ground-based ASAS-3 data. In our frequency analysis, we detected a rotational period in agreement with the previous value and nine additional harmonics. Interestingly, the amplitude of the first harmonic is higher than the fundamental rotational frequency, indicating the presence of multiple features located on diametrically opposite sides of the star's visible surface. We performed a parabolic fit to the peak of each light maximum identified in all \tess\, sectors to obtain the accurate timings and finally fitted the linear ephemeris for this star. The updated linear ephemeris is:

\begin{equation}
\mathrm{BJD_{max}}=2458570.9785(3) +  1.6279247(7) \times E,
\end{equation}
where BJD$_{\rm max}$ is the time of light maxima at epoch E. 

\begin{table}
\centering
\begin{minipage}{\linewidth}
\caption{Frequency, amplitude and S/N from the combined light curves of all six sectors. \frot\ is the rotational frequency, and the harmonics are given as integer multiples of this frequency. The labels `f$_{n*}$', n = 1, 2 and 3 indicate the frequencies of the contaminating star.}
\label{Tab:Table1}
\end{minipage}
\bigskip
\begin{tabular}{cccc}
\hline
  &\textbf{Frequency}  &\textbf{Amplitude} & \textbf{S/N} \\
  &(d$^{-1}$)  &(mmag) &  \\
\hline

 2\frot & 1.2285581(3) & 17.733(6) & 545.2 \\
  \frot & 0.6142790(1) &  8.520(6) & 140.8 \\
 4\frot & 2.457116(3)  &  0.762(6) &  54.8 \\
 3\frot & 1.842837(8)  &  0.281(6) &  15.0 \\
 5\frot & 3.07140(2)   &  0.122(6) &  10.1 \\
 7\frot & 4.29995(2)   &  0.099(6) &  11.7 \\
10\frot & 6.14279(3)   &  0.085(6) &  11.7 \\
12\frot & 7.37135(4)   &  0.060(6) &   9.1 \\
 8\frot & 4.91423(4)   &  0.059(6) &   7.3 \\
14\frot & 8.59991(4)   &  0.052(6) &   8.9 \\
\hline
f$_{1*}$&15.80501(3)   &  0.087(6) &  15.6 \\
f$_{2*}$&14.32267(6)   &  0.038(6) &   6.5 \\
f$_{3*}$&16.98574(7)   &  0.034(6) &   5.7 \\
\hline
\end{tabular}
\end{table}

\subsubsection{TRAPPIST-South}
We conducted ground-based follow-up observations in the field surrounding HD\,100357 using the TRAPPIST-South telescope. It is a 60-cm Ritchey-Chretien telescope located at ESO-La Silla Observatory in Chile. It is equipped with a thermoelectrically cooled 2K$\times$2K FLI Proline CCD camera with a field of view of $22\arcmin\times22\arcmin$ and a pixel scale of 0.65\arcsec/pixel \citep{2011Msngr.145....2J,Gillon2011}. The observations were carried out on 18 March 2025 over a duration of four hours, with individual exposures of 10\,s, through the standard Johnson--Cousins $R$-band filter. We used the python package \texttt{Prose}\footnote{\url{https://prose.readthedocs.io/en/latest/ipynb/sources.html}} \citep{Garcia2022PROSE} for analyzing the images and subsequently performing differential photometry. The primary aim of these observations was to assess the possible  contamination in the TESS light curve of HD\,100357.

\subsubsection{Contamination}\label{contamination}

After fitting and subtracting the rotational frequency and its harmonics from the time-series data, the residuals show low-amplitude frequencies at 15.805, 14.323, and 16.986\,\cd\ (see inset of middle panel of Fig.\ref{fig:lc_fourier}). Since the \tess\, pixels have poor spatial resolution (21 arcsec), the contamination of signals from nearby bright stars is expected. To mitigate this, we performed a pixel-by-pixel investigation of the \tess\, target pixel file to check for any contamination. We used the technique described in \citet{10.1093/mnras/staf361}, by selecting custom apertures around the target star identified through Gaia DR3 positions \citep{2023A&A...674A...1G} (See Fig.\,\ref{fig:combined}). Our finding is that the additional frequencies detected in the residuals of HD\,100357 (Gmag = 8.98) originated from a nearby star (Gaia DR3 5236626819707864192; Gmag = 13.01; $\Delta$Gmag = 4.03), possibly a new $\delta$ Scuti star. We further confirmed the origin of these frequencies with ground-based observations from the TRAPPIST-South telescope, which has a spatial resolution 30 times better than that of \tess\,, allowing the contaminating star to be fully resolved in the corresponding CCD frames. Fig.~\ref{fig:ground} shows the folded light curve using the dominant contaminating frequency of 15.8054 \cd, along with a CCD image of the field that marks the stars. Apart from this, there is a very close visual companion to the star that can cause contamination, which will be discussed in the following section.

\subsection{GAIA Astrometry}\label{astro_gaia}
HD\,100357 was classified as a visual double star in the Washington Double Star Catalogue (WDS) \citep{2001AJ....122.3466M} and the Tycho Double Star Catalogue \citep{2002A&A...384..180F}. Gaia DR2 \citep{2018A&A...616A...1G} and DR3 \citep{2023A&A...674A...1G} catalogues also reported this companion star. From the Gaia DR3 position and parallax, they are separated by $\sim$2.55 arcsecs in the sky and $\sim$5.33 pc apart in space. \citet{2021ApJS..254...42B} and \citet{2022A&A...657A...7K} searched for proper motion anomalies in several stars using the Hipparcos and Gaia proper motions and parallax, but they did not report any significant proper motion anomaly for HD\,100357. Hence, we can conclude that these are comoving stars with no orbital interaction. Nevertheless, the nearby visual companion with a \textit{Gaia} $G$ magnitude of 12.35 compared to the target's $G = 8.98$, may cause flux contamination in the \tess\, pixels at a level of approximately $5.7$\,\%. Since we did not detect any additional variability from this nearby companion, it will only add a constant noise level to the signal from HD\,100357; hence, one can neglect the contamination in this case. 

\subsection{Spectroscopy}
We acquired 20 high-resolution spectra for the spectroscopic analysis of HD\,100357. One spectrum was taken with the High Resolution \'Echelle Spectrograph (HRS) \citep{2012PhDT........40T} attached to the 10.0-m Southern African Large Telescope (SALT) of the South African Astronomical Observatory (Sutherland, South Africa) with spectral resolution R $\approx$ 37,000 over the wavelength range 370--890\,nm. We retrieved one archival spectrum with a spectral resolution R$\approx$48,000 over 350--920\,nm, from the Science Portal of the European Southern Observatory (ESO)\footnote{\url{https://archive.eso.org/scienceportal/home}} that was observed with the Fiberfed Extended Range Optical Spectrograph (FEROS) \citep{1999Msngr..95....8K} at MPG/ESO 2.2-m telescope (La Silla, Chile). Finally, a dedicated program was carried out with the High Efficiency and Resolution Canterbury University Large Échelle Spectrograph (HERCULES) \citep{2002ExA....13...59H}, equipped on the 1-m McLellan telescope of Mount John University Observatory (Mount John, New Zealand). In total, 18 HERCULES spectra were collected with a spectral resolution of R$\approx$41,000 over the wavelength range 380--880\,nm. The standard reduction, wavelength calibration, and continuum normalisation of HERCULES spectra were performed using the dedicated pipeline HRSP \citep{2004ASPC..310..575S}, while the HRS spectrum was reduced and wavelength calibrated using the SALT HRS pipeline \citep{2010SPIE.7737E..25C, 2016MNRAS.459.3068K, 2017ASPC..510..480K}. The HRS and FEROS spectra were then continuum normalised using the \textsc{continuum} task of the Image Reduction and Analysis Facility\footnote{\url{https://iraf-community.github.io/}} (IRAF). For the in-depth spectroscopic analysis, all these spectra were corrected for the barycentric motion. We calculated the S/N for each spectrum at a wavelength of 5500 \AA. The log of spectroscopic observations is given in Table\,\ref{spec}. 

\begin{table*}
    \centering
    \caption{Log of spectroscopic observations of HD\,100357 in the chronological orders. The rotational phase, radial velocity (RV) and projected rotational velocity (\vsini) are also listed. The rotational phases were calculated with reference time, t$_{0}$ = 2458570.9785 BJD.}
    \begin{tabular}{ccccccccccc}
    \hline
    Telescope&Diameter&Spectrograph&Phase &Date &BJD& Resolution&Integration& S/N& RV & vsini    \\
    &(m)&& &&(2400000+) & &(s)&(@550nm) & (\kms) & ($\pm$ 2\,\kms)    \\
    \hline
    MPG/ESO & 2.2&FEROS&0.938&08 Feb 2007 &54139.6498& 48000&800&133&-4.2 $\pm$ 0.3 & 56\\
\\
    SALT&10.0&HRS&0.334&14 Jan 2023&59958.5137&37000&300&74& -3.1 $\pm$ 0.4 & 63 \\
  \\  
    McLellan& 1.0& HERCULES&0.614&25 Dec 2023&60304.0915&41000&1800&45&2.7 $\pm$ 0.5 & 56 \\
    & & &0.628&25 Dec 2023 &60304.1135&41000&1800&51& -1.5$\pm$ 0.5 & 60\\
    & & &0.839&27 Dec 2023&60306.0830&41000&1800&18& -2.0 $\pm$ 1.0  & 63 \\
    & & &0.235&31 Dec 2023&60309.9862&41000&1800&41& -1.6 $\pm$ 0.6 & 62 \\
    & & &0.249&31 Dec 2023&60310.0082&41000&1800&34& -1.2 $\pm$ 0.5 & 61 \\
    & & &0.264&31 Dec 2023&60310.0324&41000&1800&33& -2.1 $\pm$ 0.5 & 63 \\
    & & &0.277&31 Dec 2023&60310.0544&41000&1800&45& -1.7 $\pm$ 0.4 & 63 \\
    & & &0.883&01 Jan 2024&60311.0397&41000&1800&53& -1.6 $\pm$ 0.4 & 61 \\
    & & &0.896&01 Jan 2024&60311.0617&41000&1800&53& -1.1 $\pm$ 0.4 & 61 \\
    & & &0.513&02 Jan 2024&60312.0662&41000&1800&59& -1.6 $\pm$ 0.4 & 60 \\
    & & &0.528&02 Jan 2024&60312.0896&41000&1800&44& -0.9 $\pm$ 0.4 & 60 \\
    & & &0.461&23 Jan 2024&60333.1452&41000&1800&40& -4.0 $\pm$ 0.5 & 59 \\
    & & &0.925&19 Feb 2024&60359.9473&41000&1800&30& -3.0 $\pm$ 0.6 & 58 \\
    & & &0.939&19 Feb 2024&60359.9693&41000&1800&39& -3.3 $\pm$ 0.5 & 57 \\
    & & &0.561&20 Feb 2024&60360.9817&41000&1800&67& 0.1 $\pm$ 0.4 & 60 \\
    & & &0.574&20 Feb 2024&60361.0037&41000&1800&63& 0.9 $\pm$ 0.4 & 58 \\
    & & &0.184&21 Feb 2024&60361.9966&41000&1800&65& -1.6 $\pm$ 0.4 & 61 \\
    & & &0.198&21 Feb 2024&60362.0186&41000&1800&55& -1.5 $\pm$ 0.4 & 61 \\
     
    \hline
    \end{tabular}
    \label{spec}
\end{table*}

\subsubsection{Least-squares Deconvolution}

\begin{figure}
    \includegraphics[width=\columnwidth]{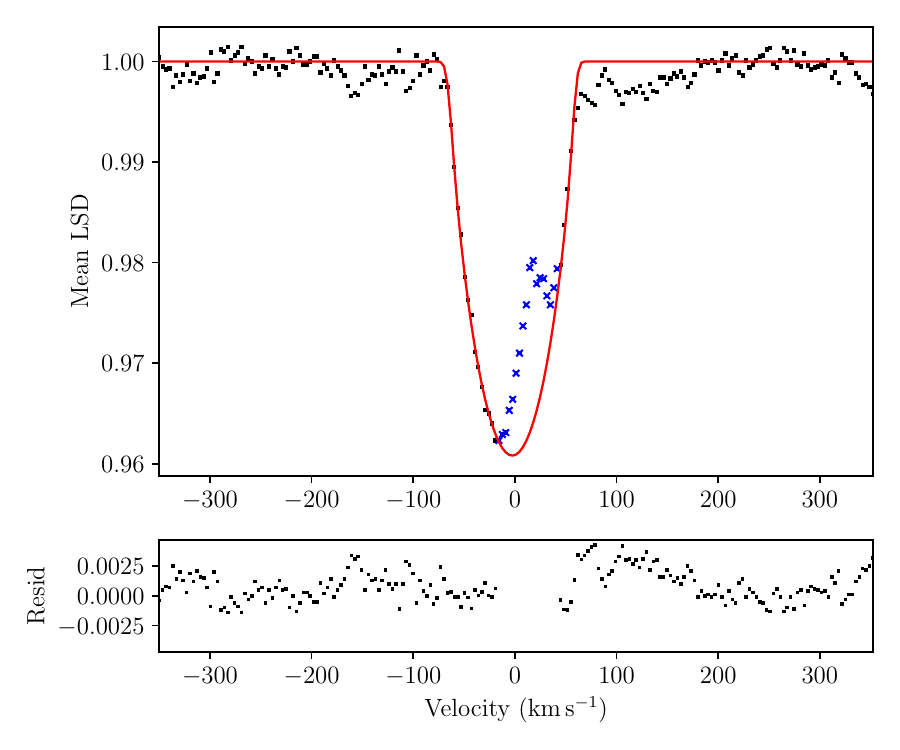}
    \caption{Top panel: The scattered points represent the LSD profile of HD\,100357 obtained from SALT spectrum. The solid curve is the best-fit rotationally broadened line profile. The points in the middle, represented by cross symbols, are excluded from the line profile fit. Bottom Panel: The residuals of the best-fit line profile to the observed LSD profile.}
    \label{fig:lsd_ex}
\end{figure}

Magnetic CP stars show significant line profile variations as a result of chemical spots on their surface. These variations are difficult to detect when the S/N of a spectrum is very low. 
To overcome this difficulty, the Least-Squares deconvolution (LSD) technique \citep{donati1997, 2010A&A...524A...5K} is used to get a higher S/N mean line profile using a large number of metallic lines. We constructed the LSD profiles for the observed spectra of HD\,100357 using a line mask prepared with the atomic data extracted from the VALD3 database \citep{1995A&AS..112..525P, 2015PhyS...90e4005R, 2019ARep...63.1010P}. We avoided the broad Balmer lines and the interstellar Na lines for constructing the LSD profiles. We fitted a rotationally broadened profile to the LSD profiles, excluding the core areas (see Fig.~\ref{fig:lsd_ex}), where line profile variations occur, to derive the radial velocities and \vsini\ values for each observation. The results derived from the LSD profiles are shown in the last two columns of Table~\ref{spec}.

\section{STELLAR PARAMETERS}\label{parameters}
\subsection{Photometric Parameters}

To estimate the effective temperature, $T_\mathrm{eff}$, and surface gravity, $\log g$, of our target, we used photometric data published in the General Catalogue of Photometric Data\footnote{\url{https://gcpd.physics.muni.cz/}} \citep{1997A&AS..124..349M} for the Str\"omgren and Geneva systems.
HD\,100357 is located close to the Galactic equator, a region having significant interstellar reddening. A colour excess of $E(B-V) = 0.10$\,mag was determined from the intensity of the Na\,\textsc{D}1 line, following the calibration by \cite{1997A&A...318..269M}, and the parallax is $\pi = 2.72 \pm 0.02$\,mas \citep{2021A&A...649A...1G}. In addition, we used the 3D dust map G-Tomo\footnote{\url{https://explore-platform.eu/sdas/about/gtomo}} \citep{2022A&A...661A.147L, 2022A&A...664A.174V} to derive the colour excess $E(B-V) = 0.136$, which is in good agreement with the value derived from the Na\,\textsc{D}1 line. 

Using the calibrations given by \cite{1993A&A...268..653N} for indices in Str\"omgren system results in the estimated $T_{uvby}=12,420$\,K and \logg\,=\,4.56. At the same time, the effective temperature calibrated using index $[u-b]$ turns out to be 1,000\,K lower: 11050\,K. It is a typical feature of CP stars with strong chemical anomalies. In light of anomalies, it sounds reasonable to apply the special relation from Eq. (12) of the original publication. This brings us $T_{[u-b]}({\rm Ap})=10750$\,K. Unfortunately, \cite{1993A&A...268..653N} did not provide the method to estimate the errors. Following the common practice, it would be reasonable to assume that the typical error of \teff\ is close to 300\,K, while for \logg\ one can expect an error of order 0.15. To account for the interstellar reddening in the Geneva data, we converted $E(B-V)$ to $E(B2-V1)$ by dividing it by 1.14.

Calibrations published by \cite{1997A&AS..122...51K} for the Geneva system, depending on the expected metallicity, give two sets of data: $T_\mathrm{Geneva}=11,726$\,K and $\log g=4.59$ for the case of $[M/H]=0$, or $T_\mathrm{Geneva}=11,560$\,K and $\log g=4.56$ when $[M/H]=+1$. The estimated error of determination is 75\,K for \teff\ and 0.07 for \logg.

As the problem of finding basic parameters from photometry is especially acute, we also revised the mentioned parameters found from the original calibrations in the framework of the study by \citet{2008A&A...491..545N}. This gives the corrected values: $T'_{uvby}=11,480$\,K and $T'_\mathrm{Geneva}=10,975$\,K.

To summarise, we can state that photometry cannot give consistent estimations of the effective temperature. We can sort all results into two groups with the mean values: $T_\mathrm{orig}=11,900$\,K and $T_\mathrm{Ap}=11,070$\,K. The first value is used as the effective temperature of HD\,100357 through this study. However, we cannot exclude that the real \teff\, can be lower by about 1,000\,K. As for the surface gravity, we adopted the mean value, $\log g=4.57\pm0.15$, which still appears overestimated.

\subsection{Distance, Luminosity, and Radius}

The distance of a star can be determined by the Gaia DR3 parallax. The parallax angle $\pi=2.72\pm0.02$\,mas results in the distance of our target $d=367\pm3$ pc. We calculated the extinction in the V band $A_{v}$ = $R_{\rm v} \times E({\rm B-V})$, where $R_{\rm v}$=3.1 \citep{1989cardelli}. We estimated the bolometric correction (BC = -0.65) using the empirical relation given by \citet{2010Torres}. For calculating the Johnson V magnitude of the star, we used the conversion from Tycho $V_{\rm T}$ and $B_{\rm T}$ magnitude \citep{2000PASP..112..961B}, resulting in the apparent magnitude, V = 8.97 $\pm$ 0.02. This yields the value of $\log\left({L/L_{\odot}}\right)$=1.86 $\pm$ 0.02 $L_{\odot}$. The radius of the star is calculated from the Stefan-Boltzmann equation, and the value is R = 2.03 $\pm$ 0.10 R$_{\odot}$. 

\subsection{Inclination Angle}
The inclination angle of the rotation axis of a magnetic CP star is required to model the abundance spot on the surface of the star and determine the geometry of the magnetic fields \citep{1950MNRAS.110..395S, 2017A&A...597A..58K}. The equatorial velocity of a rotating star can be calculated as v$_{eq}$ = 2$\pi$R/P, where R is the radius of the star and P is its rotational period. We used the rotational period determined from the \tess\ data, and mean \vsini\ ($60 \pm 2$ \kms) calculated from the spectroscopy, giving an inclination angle, $i$ = 72$^{\circ}$ $\pm$ 11$^{\circ}$. 

\subsection{Spectroscopic Parameters}\label{parm_spec}

The absorption-line spectrum of HD\,100357 aligns with that of an early A-type photosphere. Comparison with the spectral standards from \citet{2009ssc..book.....G} shows that He I lines are weak across a wide spectral range. In contrast, lines of Si\,{\sc ii}, Sc\,{\sc ii}, Ti\,{\sc ii}, Cr\,{\sc ii}, Mn\,{\sc ii}, Fe\,{\sc ii}, Co\,{\sc ii}, and Sr\,{\sc ii} are unusually strong. Notably, the Mg\,{\sc ii} 4481 \r{A} line, which is typically one of the strongest, appears relatively shallow in the spectrum of HD\,100357, which is a typical feature of He-weak stars.

The atmospheric parameters of HD\,100357 were determined using a self-consistent iterative approach based on spectral synthesis, utilising the SALT HRS spectrum. As a starting point, we adopted the effective temperature (\teff\ = 11\,900 K) and surface gravity (\logg\ = 4.57) derived from photometric data. In the subsequent step, a He-weak atmospheric model with a hydrogen abundance of N$_{H}$/N$_{tot}$ = 0.99 was computed using the ATLAS12 model atmosphere code \citep{2005MSAIS...8...25C, 2005MSAIS...8...14K} that accounts for the impact of chemical composition anomalies on the atmospheric opacity distribution.

Theoretical spectra were computed assuming local thermodynamic equilibrium (LTE) using this model atmosphere and atomic data obtained from the VALD3 database \citep{1995A&AS..112..525P, 2015PhyS...90e4005R, 2019ARep...63.1010P} with the \texttt{synthV\_NLTE} code \citep{2019ASPC..518..247T}. To compare the theoretical stellar spectra with observational data, we utilised the widget program \texttt{BinMag}\footnote{\url{http://www.astro.uu.se/~oleg/binmag.html}} \citep{2018ascl.soft05015K} that interfaces to the \texttt{synthV\_NLTE} code and allows for automatic determination of the best fits to the observed line profiles, enabling determination of highly accurate chemical abundances.

\begin{figure}
    \includegraphics[width=\columnwidth]{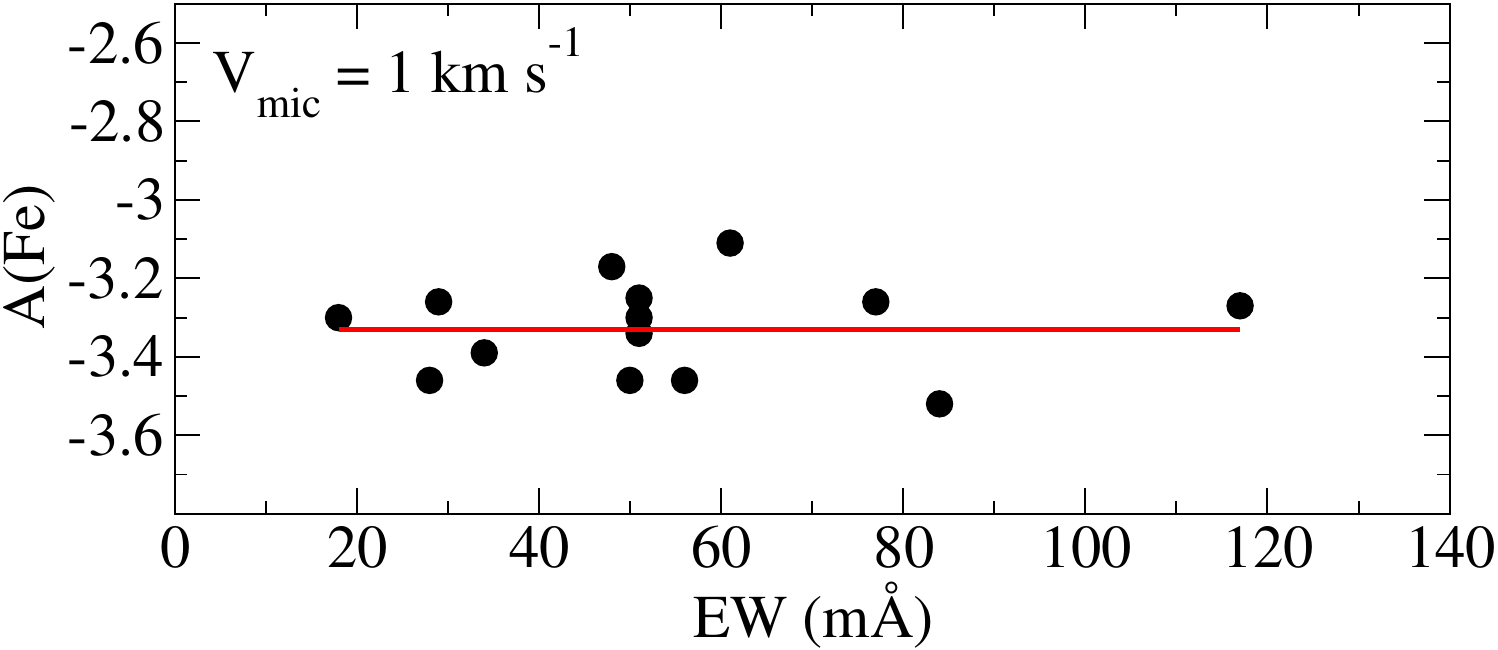}
    \caption{The scatter points represent iron abundances as a function of their equivalent widths for HD\,100357. The solid straight line is the optimal minimised slope of this relationship at \vmic\ = 1 \kms.}
    \label{fig:Fig2}
\end{figure}

\begin{figure*}
        \centering
        \begin{multicols}{2}
            \includegraphics[width=\linewidth]{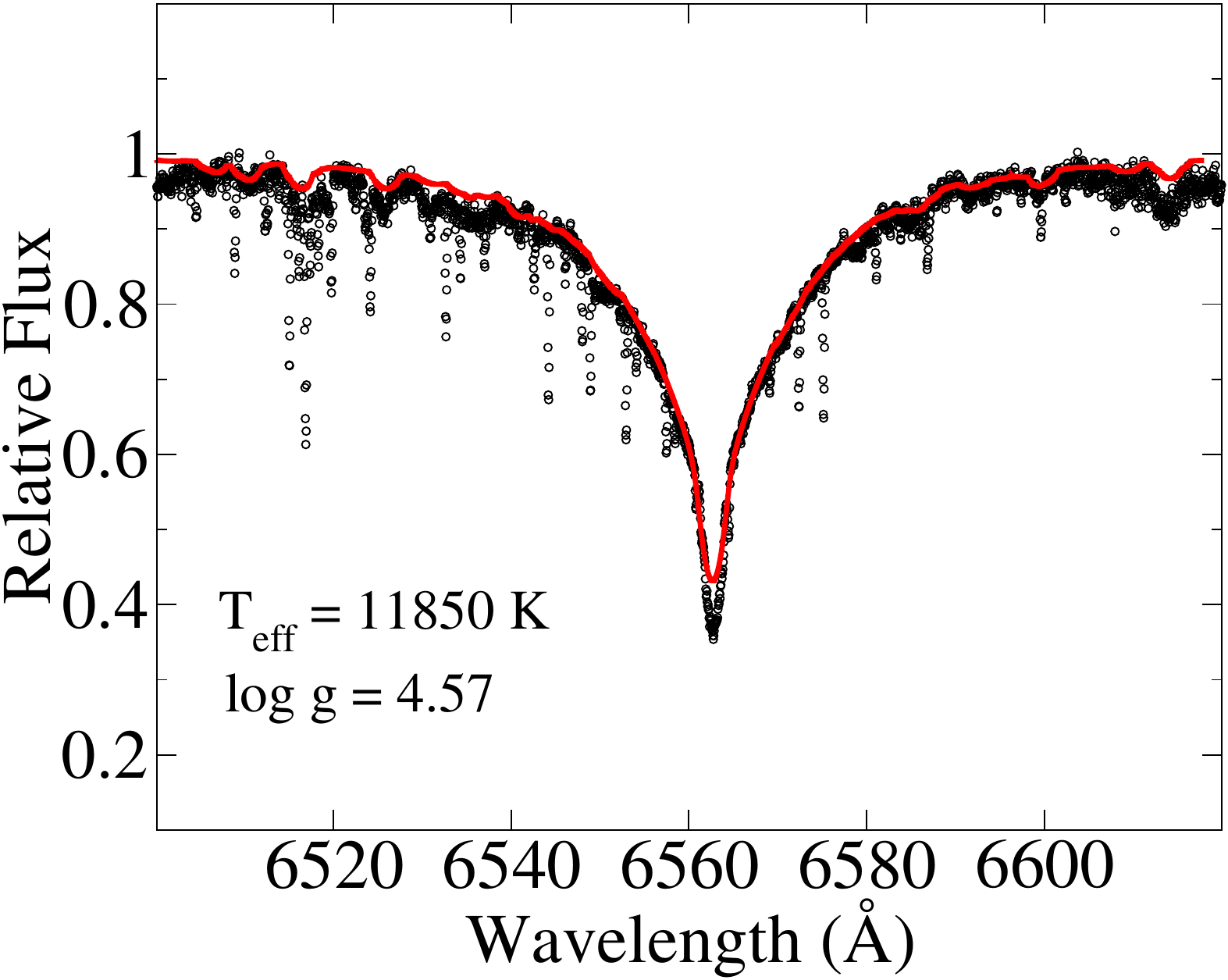}\par
            \includegraphics[width=\linewidth]{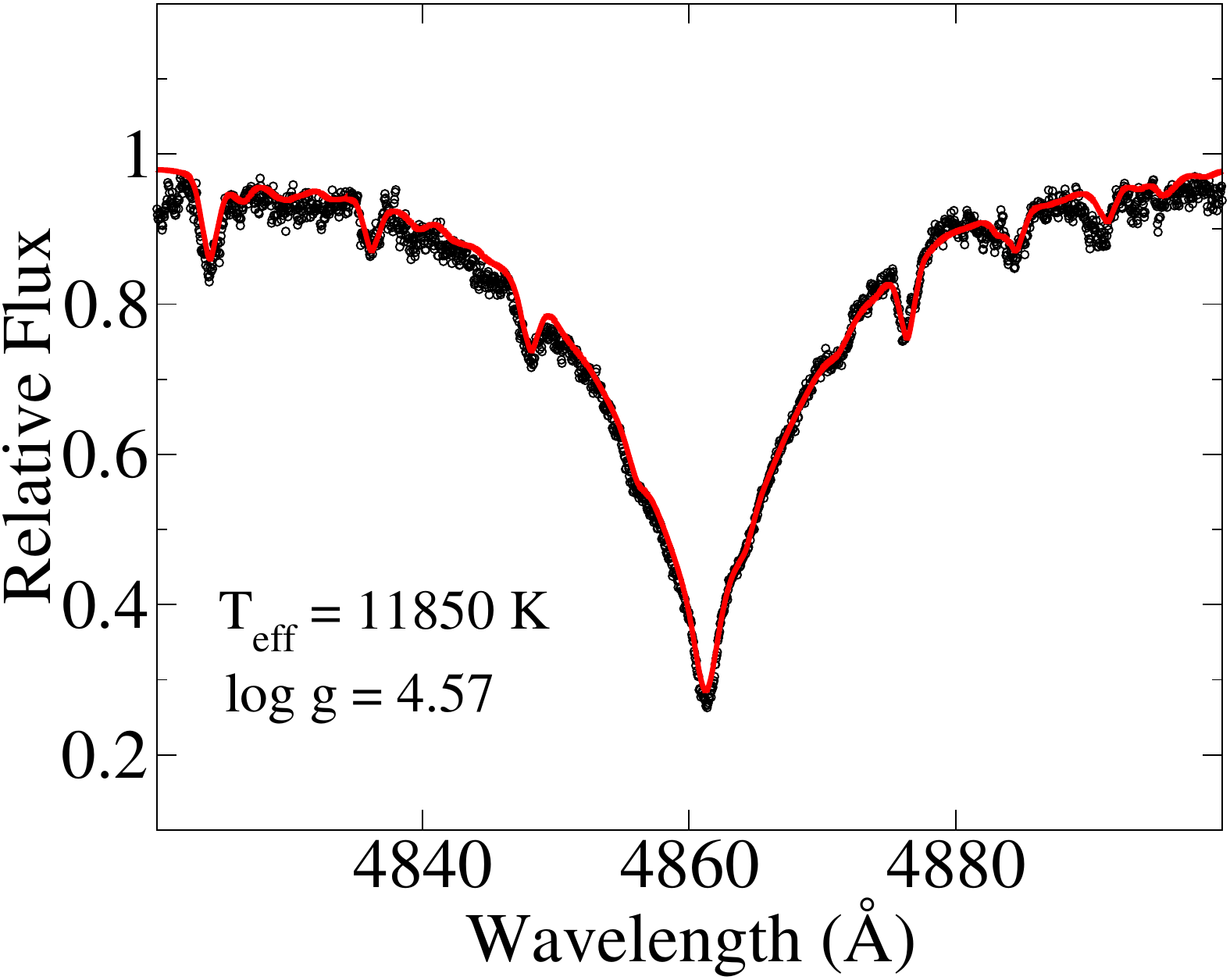}\par
        \end{multicols}
    \caption{Comparison of the observed spectrum (black circles) and  synthetic spectrum (red line) in the region of H$_\alpha$ (left panel) and H$_\beta$ (right panel). The telluric lines near the H$_\alpha$ region were excluded during the fitting.}
    \label{Balmer_lines}
    \end{figure*}

The \teff\ and \logg\ parameters were refined by comparing the synthetic spectrum to the observed spectrum in the wings of the $H_{\alpha}$ and $H_{\beta}$ hydrogen lines. Fig.~\ref{Balmer_lines} presents a comparison between the observed and synthetic profiles 
calculated using this final set of parameters. The microturbulent velocity, \vmic\ = 1 \kms, was determined using the classical method of minimising the slope in the relationship between elemental abundance and equivalent widths (see Fig.\,\ref{fig:Fig2}). The equivalent widths were derived using synthetic line-profiles, accounting for blending due to rotational broadening (60 \kms\,), ensuring that the contribution of each line was properly modeled.

Our analysis confirmed that the initial photometry-based parameters are consistent with the spectroscopic examination of the hydrogen lines. As a result, we determined the following parameters: effective temperature \teff\ = 11,850 $\pm$ 150 K, surface gravity \logg\ = 4.57 $\pm$ 0.05, and a microturbulent velocity \vmic\ = 1.0 $\pm$ 0.2 \kms. The iron abundance is measured using 14 lines, and the average value is log A(Fe) = -3.33 $\pm$ 0.12, which reflects an overabundance relative to solar with \feh\ = 1.16 dex \citep{2021SSRv..217...44L}. 

\subsection{Evolutionary Status}

\begin{figure}
    \includegraphics[width=\linewidth]{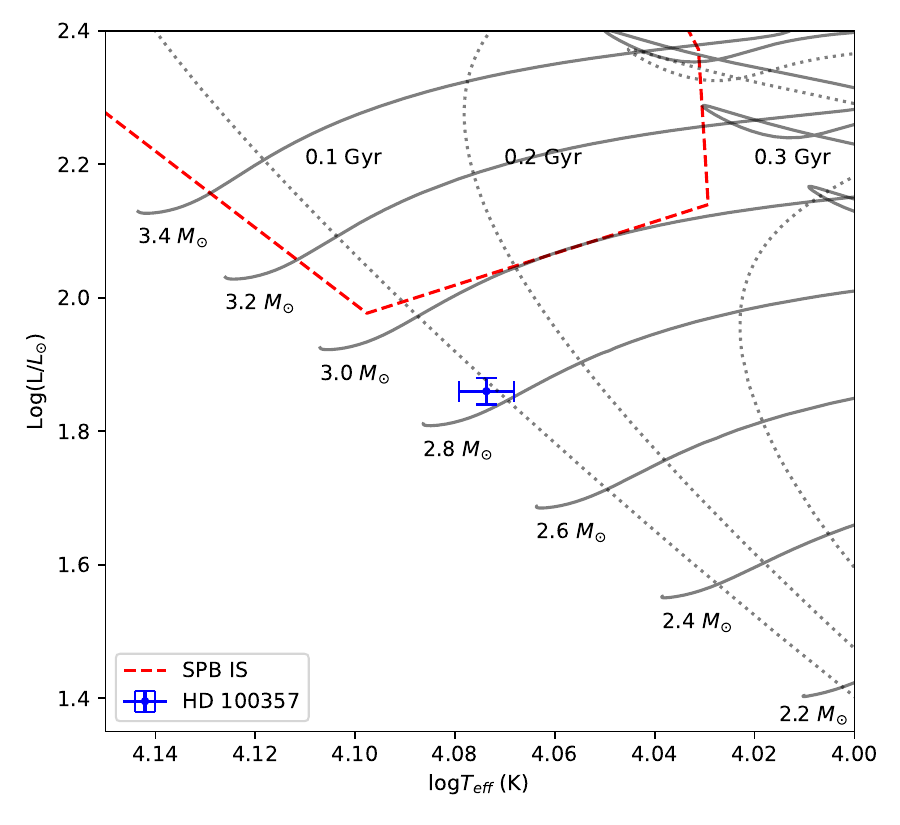}
    \caption{Position of HD\,100357 in the H-R diagram with error bars. The dotted lines and the solid lines represent the isochrones and the evolutionary tracks, respectively. The dashed line is the instability strip of slowly pulsating B stars \citep{2007CoAst.151...48M}.}
    \label{HRD}
\end{figure}

To determine the evolutionary status of the star, we constructed a Hertzsprung-Russell (H-R) diagram with theoretical isochrones and evolutionary tracks downloaded from MESA Isochrones and Stellar Tracks (MIST) \citep{2016ApJS..222....8D, 2016ApJ...823..102C} compiled using the Modules for Experiments in Stellar Astrophysics (MESA) code \citep{2011ApJS..192....3P, 2013ApJS..208....4P, 2015ApJS..220...15P, 2018ApJS..234...34P}. We placed our star in the H-R diagram using the determined values of log \teff\ and \luminosity\ with associated errors. The position of HD\,100357 in the HR diagram (Fig.~\ref{HRD}) along with the inspection of its high-resolution spectrum in Sec.~\ref{parm_spec}, indicates that it is a main-sequence star. From the inspection of the isochrones and stellar tracks, one can conclude that HD\,100357 has a mass, M $\sim$ 2.83 \mdot\ and an age, $t\sim90$ Myr. The stellar parameters determined from our analysis in this section are listed in Table~\ref{pars}.

\begin{table}
    \centering
    \caption{Summary of all stellar parameters of HD\,100357 as ascertained from the current study. The associated error bars for the last digits of the relevant values are shown in brackets.}
    \begin{tabular}{cc}
    \hline
    Parameters           & values             \\
    \hline
    P$_{\rm rot}$ (days) & 1.6279294 (7)       \\
    t$_{0}$ (BJD)        & 2458570.9743 (4)    \\
    Vmag                 &   8.97 (2)  \\
    E(B-V)               &  0.136 (2) \\
    \luminosity          &   1.86 (2)  \\
    \teff\ (K)           &  11850 (150)   \\
    \logg                &   4.57 (5)  \\
    \feh\                &   1.16 (12) \\
    \vsini\ (\kms)       &     60 (2)     \\
    \vmic\ (\kms)        &      1.0 (2)   \\
    R (R$_{\odot}$)      &   2.03 (10)  \\
    $i$ (deg)            &     72 (11)    \\
    M (\mdot)            &   2.83 (4)  \\
    Age (Gyr)            &   0.09 (4)  \\
    \hline
    \end{tabular}
    \label{pars}
\end{table}

\section{Chemical Composition}\label{abundance}

We used the SALT HRS spectrum to determine the chemical abundances of HD\,100357, which was obtained in the most recent epoch and with the best S/N. Elemental abundances were generally derived from multiple absorption lines within the observed spectral range. Abundances are expressed as \logAX\ = log (N$_{X}$ / N$_{H}$), where N$_{X}$ denotes the number density of element X and N$_{H}$ represents the number density of hydrogen. The abundances were determined under the LTE assumption by fitting synthetic line profiles to observed ones using the \texttt{BinMag} tool with the line list extracted from the VALD3 database. Hyperfine splitting \citep{2019ARep...63.1010P} was also incorporated during the extraction of the line list. The mean abundances of each element X were calculated by averaging many lines. Uncertainties were determined as the standard deviation of the mean value when more than two lines were available. Relatively high rotational broadening complicates the task of selecting lines. Consequently, most elements in the HD\,100357 spectrum are represented by a limited number of blended lines.

\begin{figure}
    \includegraphics[width=\columnwidth]{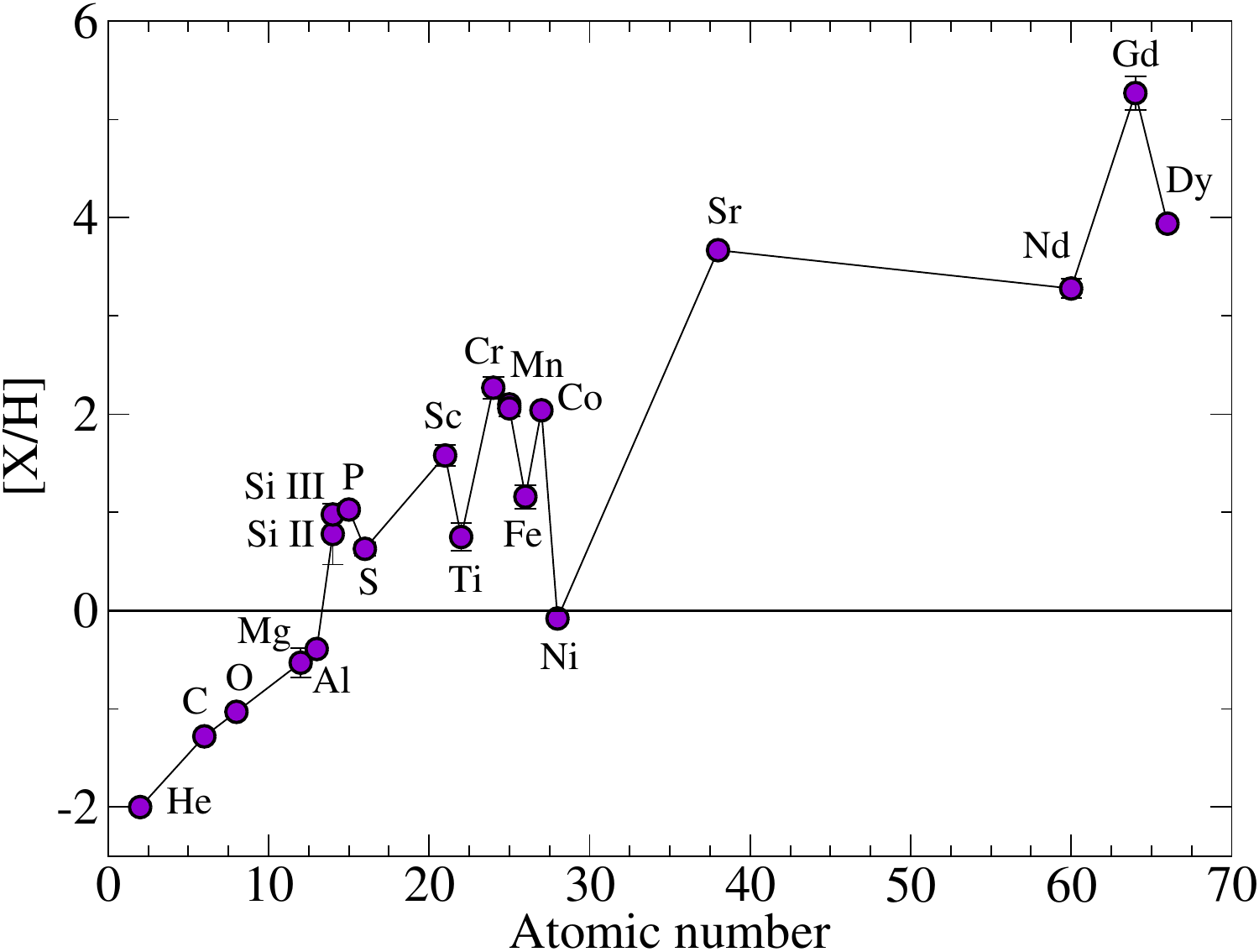}
    \caption{Relative photospheric abundance of chemical species of HD\,100357 at rotational phase 0.334. The horizontal line at zero denotes the solar reference. There is an excess of Si, Fe-peak elements, and heavier elements, accompanied by a deficiency of light elements and a pronounced underabundance of He.}
    \label{fig:Fig3}
\end{figure}

We derived the LTE abundances for 19 elements. The results of this analysis are summarized in Table\,\ref{abund_lte} and illustrated in Fig.\,\ref{fig:Fig3}, which compares the derived abundances to solar abundances using the notation [X/H] = log (N$_{X}$/N$_{H}$) - log (N$_{X}$/N$_{H}$)$_{\odot}$. The solar reference abundances are taken from \citet{2021SSRv..217...44L}. The detailed abundances from individual lines are listed in Table~\ref{det_ab_lte}.

\begin{table}
\centering
\caption{The mean LTE abundances of chemical species in HD\,100357. The parameter [X/H] = log (N$_{X}$/N$_{H}$) - log (N$_{X}$/N$_{H}$)$_{\odot}$, compares the observed abundances with solar abundances \citep{2021SSRv..217...44L}. The last column of the table lists the standard deviations for each species when more than two lines are available.}
\begin{tabular}{cccccc}
\hline
Atomic  & Ion & Number of lines & \logAX & [X/H] & Sigma \\
number&&&&&\\
\hline
2  & He\,{\sc ii}  &  2 & -3.1  & -2    & -    \\ 
6  & C\,{\sc ii}   &  1 & -4.7  & -1.28 & -    \\ 
8  & O\,{\sc i}    &  1 & -4.27 & -1.03 & -    \\ 
12 & Mg\,{\sc ii}  &  3 & -5.03 & -0.53 & 0.15 \\ 
13 & Al\,{\sc ii}  &  3 & -5.98 & -0.39 & 0.03 \\ 
14 & Si\,{\sc ii}  & 19 & -3.7  &  0.78 & 0.31 \\ 
14 & Si\,{\sc iii} &  1 & -3.5  &  0.98 & -    \\ 
15 & P\,{\sc ii}   &  2 & -5.55 &  1.03 & 0.07 \\ 
16 & S\,{\sc ii}   &  2 & -4.25 &  0.63 & 0.07 \\ 
21 & Sc\,{\sc ii}   &  3 & -7.27 &  1.58 & 0.11 \\ 
22 & Ti\,{\sc ii}  &  4 & -6.28 &  0.75 & 0.14 \\ 
24 & Cr\,{\sc ii}   & 11 & -4.11 &  2.27 & 0.11 \\ 
25 & Mn\,{\sc i}   &  1 & -4.45 &  2.1  & -    \\ 
25 & Mn\,{\sc ii}  & 11 & -4.49 &  2.06 & 0.08 \\ 
26 & Fe\,{\sc ii}  & 14 & -3.33 &  1.16 & 0.12 \\ 
27 & Co\,{\sc ii}  &  2 & -5    &  2.04 & -    \\ 
28 & Ni\,{\sc i}   &  3 & -5.8  & -0.08 & 0.01 \\ 
38 & Sr\,{\sc ii}  &  1 & -5.41 &  3.67 & -    \\ 
60 & Nd\,{\sc iii} &  2 & -7.27 &  3.28 & 0.1  \\ 
64 & Gd\,{\sc ii}  &  2 & -5.62 &  5.27 & 0.17 \\ 
66 & Dy\,{\sc ii}  &  2 & -6.94 &  3.94 & 0.02 \\ 
\hline
\end{tabular}
\label{abund_lte}
\end{table}

\subsection{Helium}

\begin{figure*}
        \centering
        \begin{multicols}{2}
            \includegraphics[width=\linewidth]{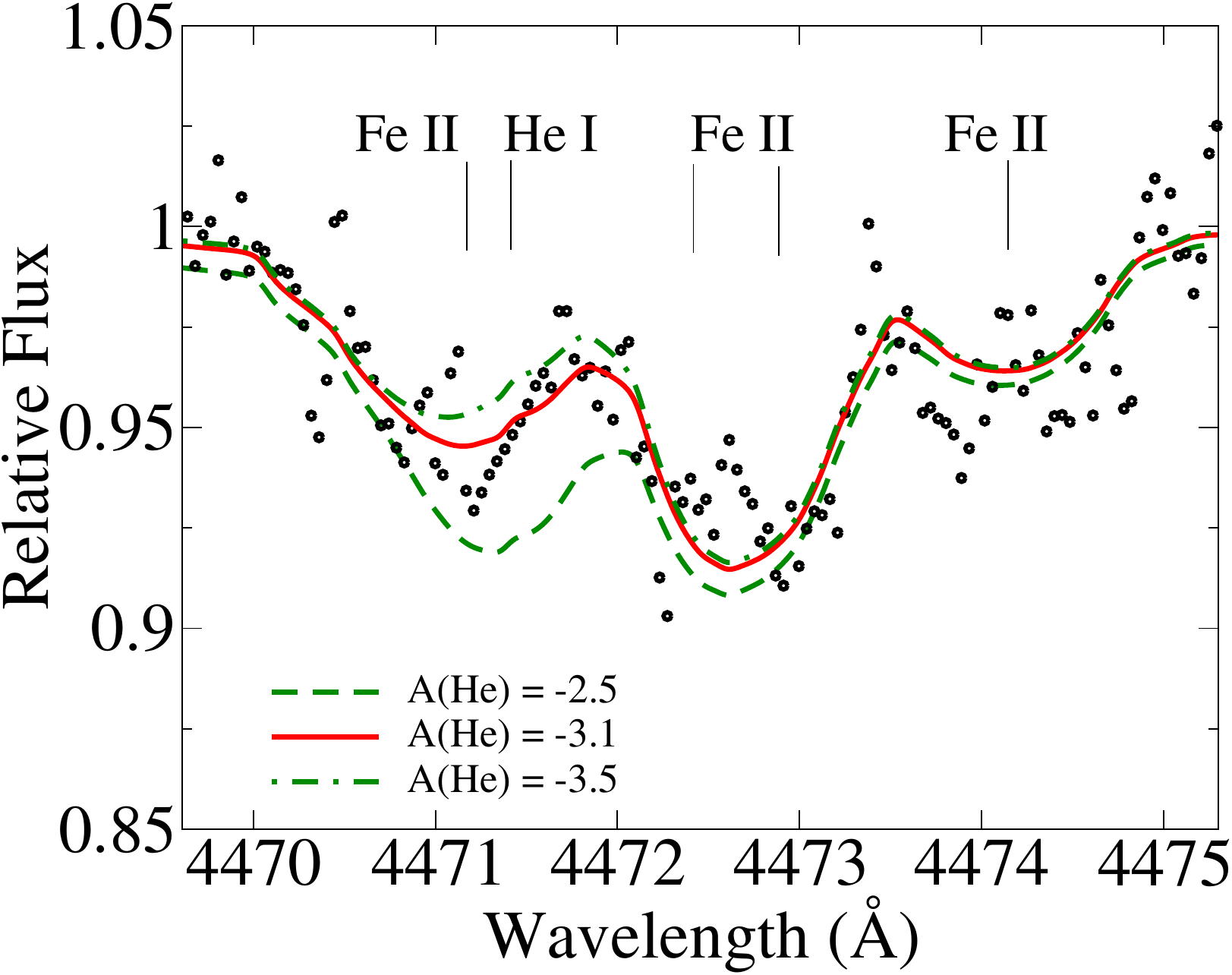}\par 
            \includegraphics[width=\linewidth]{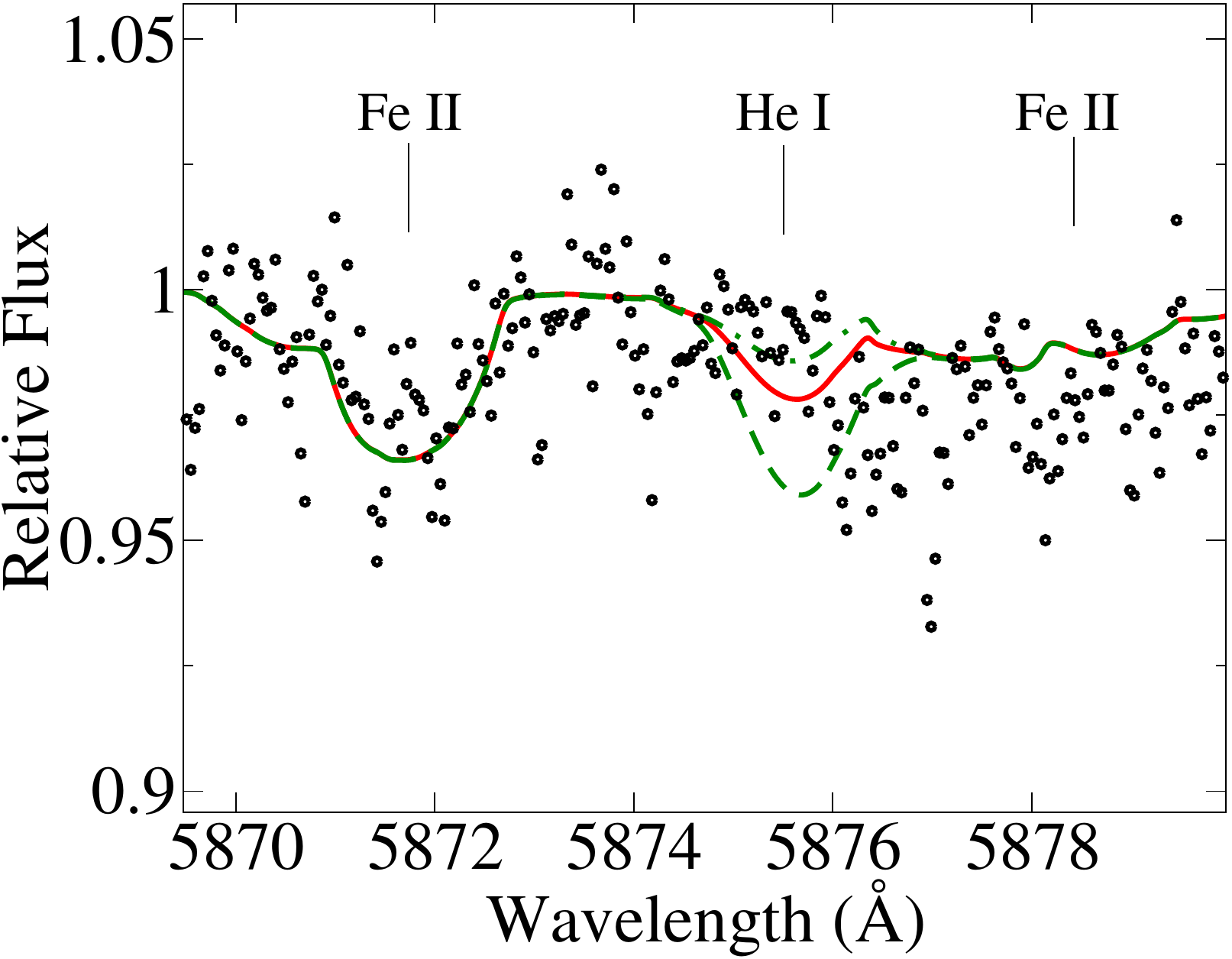}\par 
        \end{multicols}
    \caption{Observed He I $\lambda$4471 \AA\ (left panel) and He I $\lambda$5876 \AA\ (right panel) line profiles for HD\,100357 (black circles), compared to synthetic profiles (red and green lines). The red solid lines were obtained assuming the best He abundance A(He) = -3.1. The green dashed lines show the synthetic profile by increasing He to A(He) = -2.5 value. The green dashed-dot lines show the synthetic profile by decreasing He to -3.5. The nearby blended lines of Fe\,{\sc ii} are also indicated.}
    \label{fig4ab}
    \end{figure*}

We identified weak absorption lines of He\,{\sc i} at 4471 \AA\, and 5876 \AA\, and assessed the helium abundance as log(A)$_{He}$ = -3.1 dex, which is 2 dex lower than the solar value. Fig.\,\ref{fig4ab} presents a comparison between the observed and synthetic He I profiles computed using three distinct sets of He abundances for comparison.

\subsection{Carbon, Oxygen, Magnesium, and Aluminium}

We were able to measure only one C\,{\sc ii} line at 4267.259 \AA\, and the oxygen triplet in the wavelength range 7771--7775 \AA. Carbon and oxygen are both strongly depleted by more than 1.0 dex in the atmosphere of HD\,100357. Slightly less depletion was found for Mg and Al, by $\sim$0.4 dex compared to solar values.

\subsection{Silicon}

\begin{figure}
    \includegraphics[width=\columnwidth]{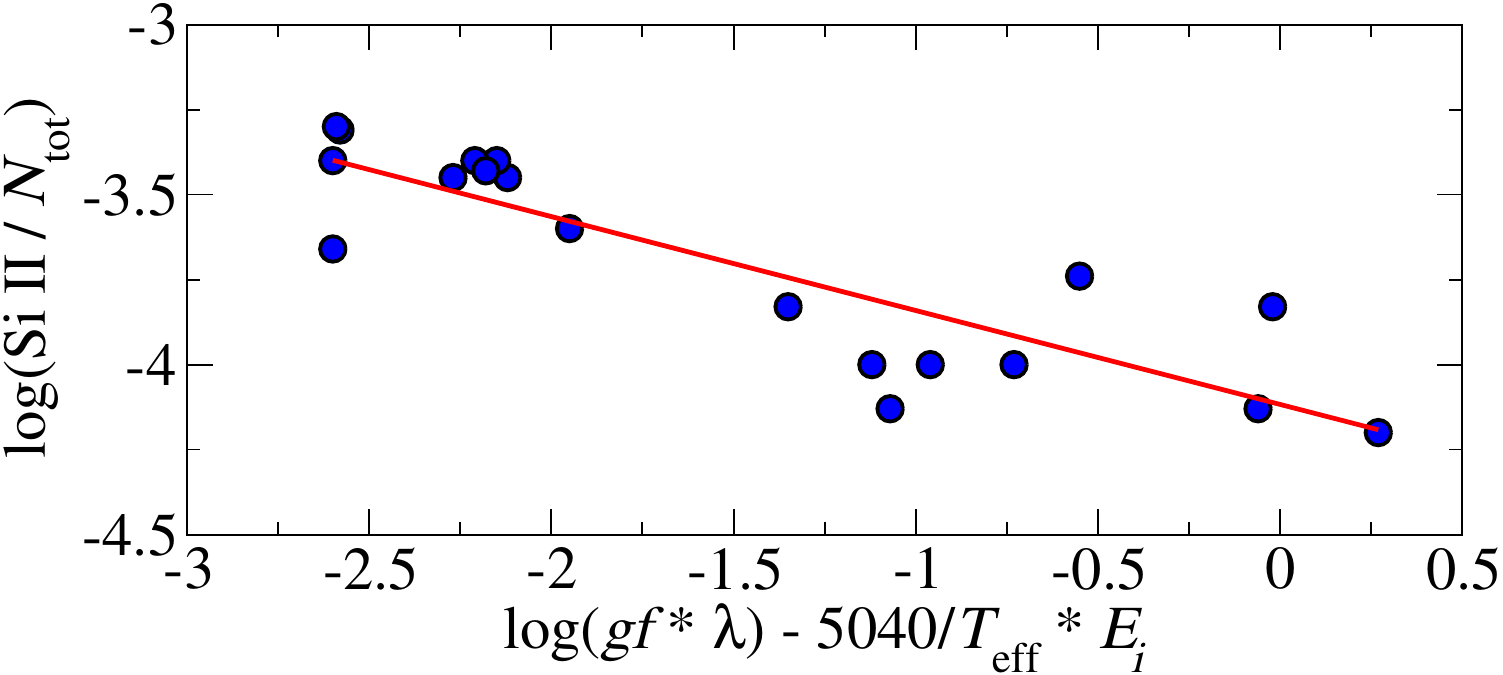}
    \caption{The scattered points indicate the individual Si abundances as a function of line strength in the atmosphere of HD\,100357. The solid line represents a linear fit to the scattered data points, illustrating the declining abundance trend of Si with line strength, indicative of abundance stratification.}
    \label{fig:Si_strat}
\end{figure}

We were able to measure only one weak Si\,{\sc iii} line, which agrees well within the error bar with the mean abundance derived from Si\,{\sc ii} lines. Despite our best efforts to select primarily unblended Si\,{\sc ii} lines with varying excitation energies and oscillator strengths, we observed a significant scatter of 0.30 dex around the mean value. The big scatter in the Si II-based abundances prompted us to investigate the relationship between abundance measurements from individual lines and the line strength, which depends on oscillator strength and excitation energy (see, e.g., \citet{2014psce.conf..220R}). This relationship is illustrated in Fig.~\ref{fig:Si_strat} for Si\,{\sc ii} lines. The figure shows a noticeable trend: stronger lines formed higher in the atmosphere yield lower individual abundances compared to weaker lines formed closer to the photosphere. These results suggest a clear vertical stratification of Si, where most of the Si is concentrated in the deeper layers of the star \citep{2013A&A...551A..30B}. A detailed analysis of vertical abundance stratification is beyond the scope of this paper.

\subsection{Phosphorous, Sulphur}

We used two lines of P\,{\sc ii} at 4602 \AA\, and 6166 \AA\, and two lines of S\,{\sc ii} at 5454 \AA, and 5606 \AA. Our estimates indicate that both P and S are overabundant, with the abundance for P being 1.03 dex higher than the solar value and a slightly lesser abundance for S by 0.63 dex.

Among the two categories of He-weak stars~--- magnetic and non-magnetic~--- non-magnetic He-weak stars occasionally show overabundances of phosphorus (up to twice the solar value) and gallium (up to five times the solar value). While phosphorus is overabundant in the atmosphere of HD\,100357, no gallium lines were detected.

\subsection{Iron peak elements: Scandium to Nickel}

With the exception of nickel, which possesses an abundance close to the solar value, the remaining six iron-peak elements exhibit significant overabundances, ranging from 1.0 to 2.0 dex.

\subsection{Neutron-capture elements}
The abundance of Sr was measured using the Sr\,{\sc ii} $\lambda$ 4215.5\,\AA\ line and shows a significant enhancement of 3.67 dex compared to the solar value. Efforts to analyse other neutron capture elements, such as Y and Zr, were unsuccessful due to a lack of suitable spectral lines for accurate abundance determination.

\subsection{Rare Earth Elements (REE)}
HD\,100357 displays significant overabundances of rare earth elements, including Nd, Gd, and Dy. Such high overabundances are a distinctive feature of upper main-sequence magnetic chemically peculiar stars. A similar chemical pattern was observed in the slowly rotating, strongly magnetic Ap star HD\,144897, as reported by \citet{2006A&A...456..329R}. Owing to the high rotational velocity of HD\,100357, we were able to reliably measure only 3 REE.

\section{Doppler Imaging}\label{Doppler}

DI \citep{2017A&A...597A..58K} is a robust technique to ascertain the topology of chemical spots on Ap stars. The technique involves mapping the surface inhomogeneities of a star given the observed line profile variations. For this, one needs high S/N (S/N$\sim$300) line profiles of different chemical elements over adequate rotational phases. In the absence of sufficient S/N, we can use the LSD technique to produce a mean line profile with enhanced S/N appropriate for DI, where the S/N is increased proportional to the square root of the number of effective absorption lines.

For our study, we computed the LSD profiles for individual elements across the various observed phases for HD\,100357. Out of the various sets of LSD profiles calculated for HD\,100357, those derived for Cr and Fe exhibit the highest S/N, enabling us to map the surface spots of these elements using DI \citep{kochukhov2016}. To this end, we employed the DI code {\tt InversLSD} \citep{kochukhov2014} to map the distributions of Cr and Fe in terms of the local line strength of the LSD profile. Following the approach of \citet{freour2023} and \citet{semenko2024}, the local spectra were computed under the assumption of a Milne-Eddington atmosphere, adopting a Voigt absorption profile. Limb darkening was treated according to the square-root law \citep{diaz-cordoves1992}. The radial velocity offset and the projected rotational velocity were refined during the DI analysis by identifying the values that yielded the best fit to the observations.

\begin{figure}
    \includegraphics[width=\columnwidth]{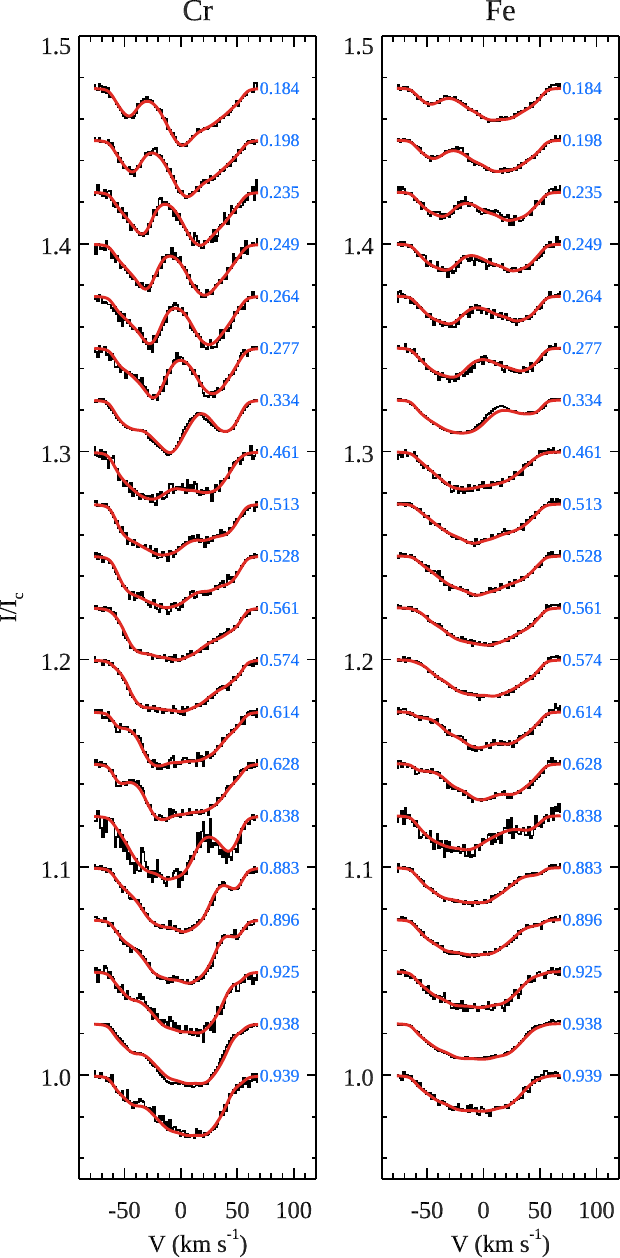}
    \caption{Comparison of the observed LSD profiles (histograms) and DI model LSD profiles (solid lines) of Cr and Fe. The profiles for different rotational phases are offset vertically. The phase values are shown to the right of the respective profiles.}
    \label{fig:di_lsd}
\end{figure}

\begin{figure*}
    \includegraphics[width=\linewidth]{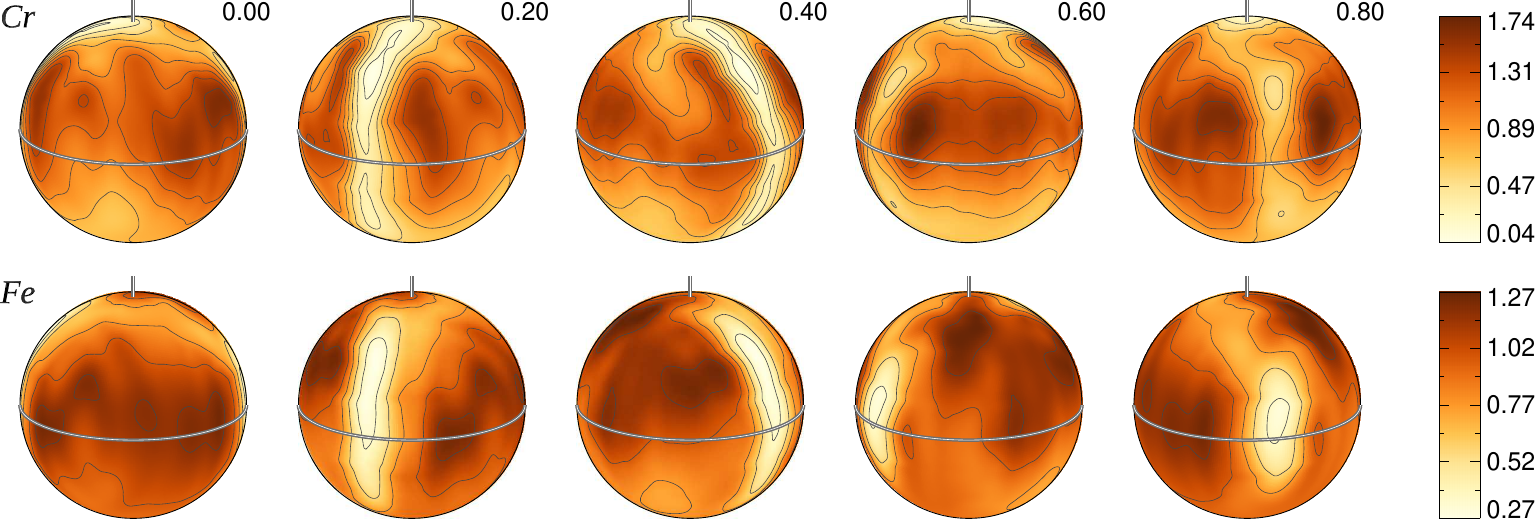}
    \caption{Surface distributions of Cr and Fe reconstructed with DI. The maps are given in terms of the logarithm of the local line strength. The star is shown at five rotational phases, indicated to the right of each column, at the inclination angle $i=72\degr$. In each spherical plot, the thick line shows the stellar equator, whereas the short bar corresponds to the rotational pole. We can notice a ring of underabundance regions inclined at $90\degr$ to the stellar equator.}
    \label{fig:di_maps}
\end{figure*}

The results of the DI calculations are presented in Fig.~\ref{fig:di_lsd} and Fig.~\ref{fig:di_maps}, which show the fits to the Cr and Fe LSD profiles and the corresponding surface line strength maps, respectively. This analysis yields \vsini\ values of 59.1 \kms\ and 59.8 \kms\ for Cr and Fe, respectively, consistent with other determinations in this study. The maps shown in Fig.~\ref{fig:di_maps} indicate that the two elements have qualitatively similar distributions, with Cr exhibiting a higher contrast surface structure. For both elements, the most prominent feature is a ring of relative underabundance encircling the star. 

\begin{figure}
    \includegraphics[width=\columnwidth]{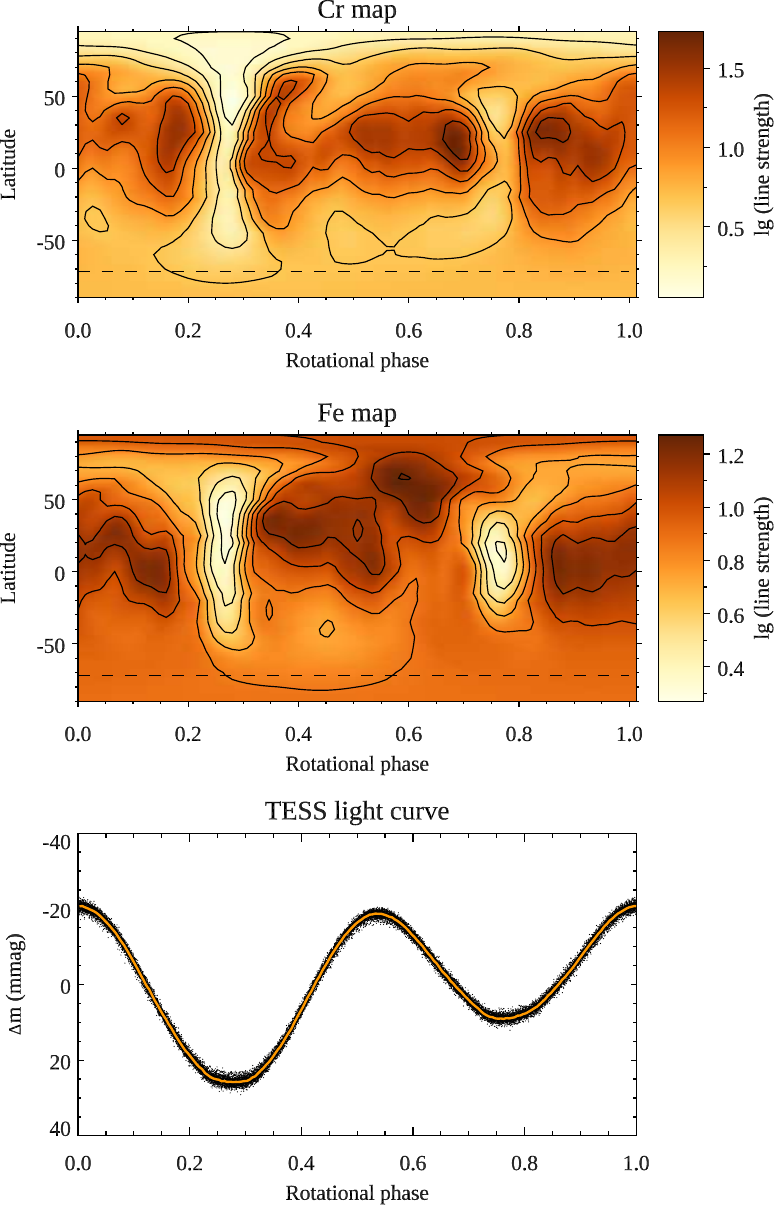}
    \caption{The top two panels show the surface distribution of Cr and Fe, respectively, as a 2D plane, where the x-axis gives the rotational phase and the y-axis represent the latitude. The dashed line represents the -72$^{\circ}$ latitude below which the surface of the star is not visible to the line of sight. At the bottom panel, the phased TESS light curve of HD\,100357 is plotted for comparison.}
    \label{fig:di_lc}
\end{figure}

Fig.~\ref{fig:di_lc} shows the same Cr and Fe maps as in Fig.~\ref{fig:di_maps}, but now in a rectangular format with the horizontal axis following the increasing rotational phase. Below the map is the phased \tess\ light curve. It is evident that the areas of relative underabundance for both elements coincide with the light minima. The result agrees with qualitative expectations from theory: a higher abundance increases UV absorption, leading to flux redistribution from UV to optical and NIR \citep{2015A&A...576A..82K}. Therefore, in the \tess\ bandpass, the star is brighter (fainter) where the element abundance is higher (lower).

\section{Discussion and Conclusions}\label{conclusion}

We report the detailed study of the Ap star HD\,100357, which exhibits an abundance pattern consistent with the typical trends observed in other hot Ap stars. The star displays slight deficiencies in several light elements. The Fe-peak elements are overabundant by about one dex, except for Cr, Mn, and Co, those are overabundant by approximately two dex. The neutron capture element Sr is significantly enhanced, with an abundance exceeding the solar value by more than 3 dex. Additionally, the rare earth elements are markedly overabundant, ranging from three to five dex above their solar values. Also, we found evidence of atomic diffusion through the abundance stratification of Si at different depths. The overabundance of weaker lines of Si indicates a dominance of gravitational settling over radiative levitation.

The Ap stars are distinguished by their excess of Fe-peak elements and rare earth elements. A notable characteristic of hot (\teff\ > 10,000\,K) counterparts in this class is an overabundance of Si. The deficiency of He is also seen in these stars, where the underabundance of He is correlated with the \logg\ of the star (see Fig.~4 of \citealt{10.1093/mnras/sty1912}). Similarly, the Fe abundance is correlated to the \teff\ of the star (see Fig.~5 of \citealt{10.1093/mnras/sty1912}). Thus, our target HD\,100357 aligns well with all the characteristics of a hot magnetic CP star.

DI analysis revealed a ring of underabundance of Fe and Cr encircling the star. Analogous to the DI results for other Ap stars \citep{kochukhov2019, kochukhov2022, kochukhov2023}, these features presumably correspond to the equatorial region of the magnetic axis of a rather weak dipolar-like magnetic field, which is inclined at about 90$^{\circ}$ to the stellar rotational axis.

Our quantitative examination of individual abundances and qualitative topological abundance distributions indicates that the star is likely to exhibit a detectable magnetic field. Consequently, HD\,100357 can be classified as a hot Ap star exhibiting anomalies in Si, Sc, Ti, Cr, Mn, Fe, Co, Sr, and He, or CP2 as per Preston's classification \citep{preston1974}. The study of these stars aids in comprehending the influence of magnetic fields and rotational forces on stellar atmospheres and chemical composition. Therefore, detailed spectropolarimetric observations are proposed to study the magnetic field topology of HD\,100357.

\section*{Acknowledgements}
We are grateful to the Indian and Belgian funding agencies DST (DST/INT/Belg/P-09/2017) and BELSPO (BL/33/IN12) for providing financial support to carry out this work in collaboration. AD acknowledges the financial support received from the DST-INSPIRE Fellowship Programme (DST/INSPIRE Fellowship/2020/IF200245). SJ acknowledges the financial support received from the BRICS grant under the project DST/IC/BRICS/2023/5. S.A. acknowledges support from the National Natural Science Foundation of China, grant No. W2432009. OK acknowledges support by the Swedish Research Council (grant agreement 2023-03667) and the Swedish National Space Agency. OT acknowledges financial support from the SEISMIC project, the Max Planck–Humboldt Research Unit in collaboration with the Max Planck Institutes for Astrophysics (MPA) and Solar System Research (MPS), and Kyambogo University, as well as from the International Science Programme (ISP) at Uppsala University. This paper includes data collected by the TESS and SIMBAD databases operated by the NASA Explorer Program and CDS, Strasbourg, France, respectively. The research leading to these results has received funding from the ARC grant for Concerted Research Actions, financed by the Wallonia-Brussels Federation. TRAPPIST is funded by the Belgian Fund for Scientific Research (Fond National de la Recherche Scientifique, FNRS) under the grant PDR T.0120.21. MG and EJ are FNRS-F.R.S. Research Directors. Funding for KB was provided by the European Union (ERC AdG SUBSTELLAR, GA 101054354). We acknowledge the use of the HERCULES spectrograph at the Mt John University Observatory, operated by the University of Canterbury, New Zealand. We acknowledge the use of the High Resolution Spectrograph (HRS) on the Southern African Large Telescope (SALT). We acknowledge Dr. Donald Wayne Kurtz for his insights into selecting this peculiar candidate from the N-C survey. We express our gratitude to Dr. Gautier Mathys for his meticulous assessment and insightful comments regarding this manuscript. 
\section*{Data Availability}

The data underlying this article will be shared upon request to
the corresponding author. The \textit{TESS} time-series flux data for the target star is publicly available from the NASA MAST archive (\url{https://mast.stsci.edu/portal/Mashup/Clients/Mast/Portal.html}). This work has made use of data from the European Space Agency (ESA) mission \textit{Gaia} (\url{https://www.cosmos.esa.int/gaia}), processed by the Gaia Data Processing and Analysis Consortium (DPAC, \url{https://www.cosmos.esa.int/web/gaia/dpac/consortium}). Funding for DPAC has been provided by national institutions, in particular those participating in the \textit{Gaia} Multilateral Agreement. This research has also made use of the SIMBAD database, operated at CDS, Strasbourg, France. This research has used data, tools, or materials developed as part of the EXPLORE project, which has received funding from the European Union’s Horizon 2020 research and innovation programme under grant agreement No.~101004214.



\bibliographystyle{mnras}
\bibliography{example} 




\appendix
\section{Individual line abundances}
\begin{table}
\centering
\caption{The LTE abundances of HD\,100357 derived from individual spectral lines of chemical species. The column [X/H] represents the abundance \logAX\, relative to the solar abundances.}
\begin{tabular}{cccccc}
\hline
Atomic & Element & Wavelength & \logAX & [X/H] & solar  \\ 
Number&& (\AA) &&&abundance\\
\hline
 2 & He II  & 5875 & -3.1 & -2 & -1.1 \\ 
 2 & He II  & 4471 & -3.1 & -2 & -1.1 \\ 
 6 & C II   & 4267.259 & -4.7 & -1.28 & -3.42 \\ 
 8 & O I    & 7771-7775 & -4.27 & -1.03 & -3.24 \\ 
12 & Mg II  & 4481 & -5.2 & -0.7 & -4.5 \\ 
12 & Mg II  & 7877.054 & -5 & -0.5 & -4.5 \\ 
12 & Mg II  & 7896.366 & -4.9 & -0.4 & -4.5 \\ 
13 & Al II  & 4663.046 & -5.94 & -0.35 & -5.59 \\ 
13 & Al II  & 5593.2998 & -6 & -0.41 & -5.59 \\ 
13 & Al II  & 7042.083 & -6 & -0.41 & -5.59 \\ 
14 & Si II  & 4130.872 & -3.83 & 0.65 & -4.48 \\ 
14 & Si II  & 4130.894 & -3.83 & 0.65 & -4.48 \\ 
14 & Si II  & 4621.418 & -3.45 & 1.03 & -4.48 \\ 
14 & Si II  & 4621.7222 & -3.45 & 1.03 & -4.48 \\ 
14 & Si II  & 5041.0239 & -3.74 & 0.74 & -4.48 \\ 
14 & Si II  & 5055.9839 & -4.13 & 0.35 & -4.48 \\ 
14 & Si II  & 5056.3169 & -4.13 & 0.35 & -4.48 \\ 
14 & Si II  & 5688.817 & -3.4 & 1.08 & -4.48 \\ 
14 & Si II  & 5800.454 & -3.31 & 1.17 & -4.48 \\ 
14 & Si II  & 5806.731 & -3.66 & 0.82 & -4.48 \\ 
14 & Si II  & 5867.48 & -3.4 & 1.08 & -4.48 \\ 
14 & Si II  & 5868.4438 & -3.4 & 1.08 & -4.48 \\ 
14 & Si II  & 5957.5591 & -4 & 0.48 & -4.48 \\ 
14 & Si II  & 6371.3711 & -4.2 & 0.28 & -4.48 \\ 
14 & Si II  & 6660.532 & -3.43 & 1.05 & -4.48 \\ 
14 & Si II  & 6671.841 & -3.6 & 0.88 & -4.48 \\ 
14 & Si II  & 6699.431 & -3.3 & 1.18 & -4.48 \\ 
14 & Si II  & 7848.816 & -4 & 0.48 & -4.48 \\ 
14 & Si II  & 7849.722 & -4 & 0.48 & -4.48 \\ 
14 & Si III & 4567.84 & -3.5 & 0.98 & -4.48 \\ 
15 & P II   & 4602 & -5.6 & 0.98 & -6.58 \\ 
15 & P II   & 6165.598 & -5.5 & 1.08 & -6.58 \\ 
16 & S II   & 5453.855 & -4.3 & 0.58 & -4.88 \\ 
16 & S II   & 5606.151 & -4.2 & 0.68 & -4.88 \\ 
21 & Sc II  & 4374.457 & -7.15 & 1.7 & -8.85 \\ 
21 & Sc II  & 4400.3892 & -7.29 & 1.56 & -8.85 \\ 
21 & Sc II  & 5526.79 & -7.36 & 1.49 & -8.85 \\ 
22 & Ti II  & 4163.644 & -6.4 & 0.63 & -7.03 \\ 
22 & Ti II  & 4443.801 & -6.2 & 0.83 & -7.03 \\ 
22 & Ti II  & 4468.493 & -6.4 & 0.63 & -7.03 \\ 
22 & Ti II  & 4571.972 & -6.12 & 0.91 & -7.03 \\ 
24 & Cr II  & 4145.7808 & -4 & 2.38 & -6.38 \\ 
24 & Cr II  & 4242.366 & -4 & 2.38 & -6.38 \\ 
24 & Cr II  & 4256.1079 & -4.23 & 2.15 & -6.38 \\ 
24 & Cr II  & 4275.5669 & -4.1 & 2.28 & -6.38 \\ 
24 & Cr II  & 4511.7749 & -4.01 & 2.37 & -6.38 \\ 
24 & Cr II  & 4539.595 & -4.15 & 2.23 & -6.38 \\ 
24 & Cr II  & 4592.052 & -4.3 & 2.08 & -6.38 \\ 
24 & Cr II  & 4616.6289 & -4.02 & 2.36 & -6.38 \\ 
24 & Cr II  & 4812.3369 & -4.25 & 2.13 & -6.38 \\ 
24 & Cr II  & 4824.127 & -4.04 & 2.34 & -6.38 \\ 
24 & Cr II  & 5420.925 & -4.06 & 2.32 & -6.38 \\ 
25 & Mn I   & 4030.7529 & -4.45 & 2.1 & -6.55 \\ 
25 & Mn II  & 4251.7168 & -4.57 & 1.98 & -6.55 \\ 
25 & Mn II  & 4252.9629 & -4.57 & 1.98 & -6.55 \\ 
25 & Mn II  & 4292.2368 & -4.43 & 2.12 & -6.55 \\ 
25 & Mn II  & 4478.6372 & -4.43 & 2.12 & -6.55 \\ 
25 & Mn II  & 4518.956 & -4.53 & 2.02 & -6.55 \\ 
\hline
\end{tabular}
\label{det_ab_lte}
\end{table}

\addtocounter{table}{-1}

\begin{table}
\centering
\caption{Continued.}
\begin{tabular}{cccccccc}
\hline
Atomic & Element & Wavelength &log (A)$_{X}$ & [X/H] & solar \\ 
Number && (\AA) &&&abundance\\
\hline
25 & Mn II  & 4727.84   & -4.6  &  1.95  & -6.55  \\ 
25 & Mn II  & 4738.29   & -4.5  &  2.05  & -6.55  \\ 
25 & Mn II  & 4749.112  & -4.51 &  2.04  & -6.55  \\ 
25 & Mn II  & 4755.727  & -4.31 &  2.24  & -6.55  \\ 
25 & Mn II  & 4764.728  & -4.45 &  2.1   & -6.55  \\ 
25 & Mn II  & 4784.625  &  -4.5 &  2.05  & -6.55  \\ 
26 & Fe II  & 4461.706  & -3.25 &  1.23  & -4.48  \\ 
26 & Fe II  & 4520.218  & -3.52 &  0.96  & -4.48  \\ 
26 & Fe II  & 4522.628  & -3.27 &  1.21  & -4.48  \\ 
26 & Fe II  & 4596.01   & -3.34 &  1.14  & -4.48  \\ 
26 & Fe II  & 4638.041  & -3.17 &  1.31  & -4.48  \\ 
26 & Fe II  & 4977.03   & -3.11 &  1.37  & -4.48  \\ 
26 & Fe II  & 5325.552  & -3.26 &  1.22  & -4.48  \\ 
26 & Fe II  & 5387.063  & -3.46 &  1.02  & -4.48  \\ 
26 & Fe II  & 5487.618  & -3.46 &  1.02  & -4.48  \\ 
26 & Fe II  & 5830.344  & -3.39 &  1.09  & -4.48  \\ 
26 & Fe II  & 5838.99   & -3.3  &  1.18  & -4.48  \\ 
26 & Fe II  & 5871.77   & -3.26 &  1.22  & -4.48  \\ 
26 & Fe II  & 6060.967  & -3.3  &  1.18  & -4.48  \\ 
26 & Fe II  & 7506.543  & -3.46 &  1.02  & -4.48  \\ 
27 & Co II  & 4569.25   & -5    &  2.04  & -7.04  \\ 
27 & Co II  & 4660.656  & -5    &  2.04  & -7.04  \\ 
28 & Ni I   & 4980.166  & -5.8  & -0.080 & -5.72  \\ 
28 & Ni I   & 5080.527  & -5.79 & -0.070 & -5.72  \\ 
28 & Ni I   & 5081.1069 & -5.81 & -0.090 & -5.72  \\ 
38 & Sr II  & 4215.519  & -5.41 &  3.67  & -9.08  \\ 
60 & Nd III & 4927.4877 & -7.34 &  3.21  & -10.55 \\ 
60 & Nd III & 5294.1133 & -7.2  &  3.35  & -10.55 \\ 
64 & Gd II  & 4514.504  & -5.74 &  5.15  & -10.89 \\ 
64 & Gd II  & 4193.145  & -5.5  &  5.39  & -10.89 \\ 
66 & Dy II  & 4573.855  & -6.92 &  3.95  & -10.87 \\ 
66 & Dy II  & 4073.12   & -6.95 &  3.92  & -10.87 \\ 
    \hline
    \end{tabular}
    \label{}
\end{table}

\section{Contamination}
\begin{figure}
  \centering
  \begin{subfigure}[b]{0.45\textwidth}
    \includegraphics[width=\textwidth]{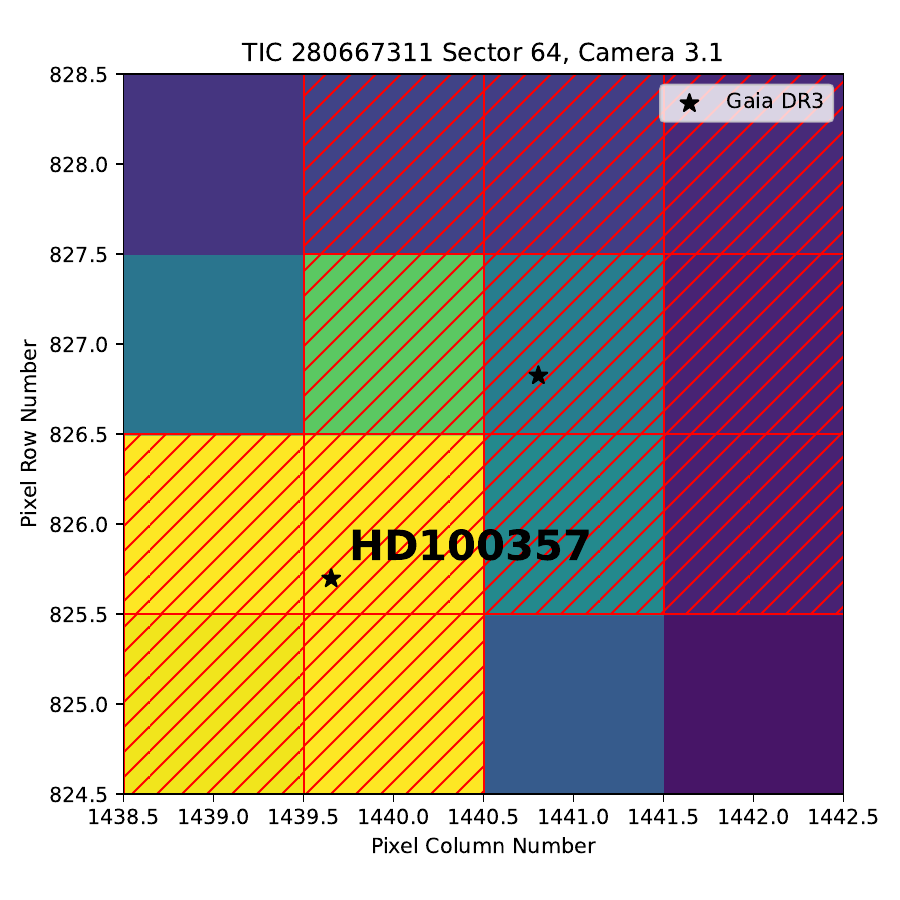}
  \end{subfigure}
  \hfill
  \begin{subfigure}[b]{0.45\textwidth}
    \includegraphics[width=\textwidth]{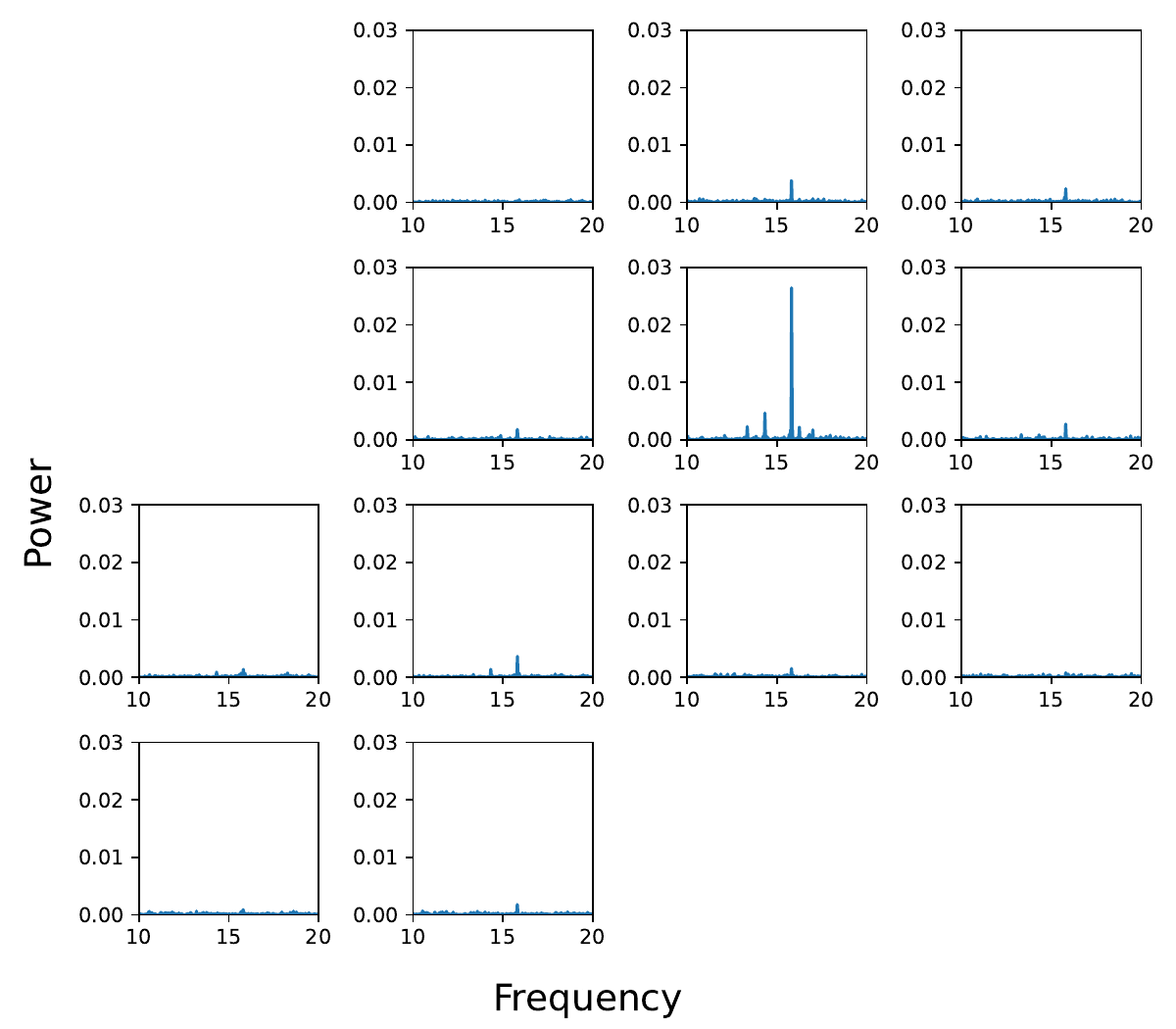}
  \end{subfigure}
  \caption{Top panel: The TESS pixels (4 $\times$ 4, Sector 10) in the field of HD\,100357 marked with a star symbol and labelled. To the top right, a nearby star is seen contaminating HD\,100357. The hatched pixels show the masks used to plot the periodograms in the bottom panel. Bottom Panel: The corresponding periodograms of the selected masks. We can see the increased amplitude of the contaminating frequency at the pixel with the contaminating star and the amplitude drops towards the target star.}
  \label{fig:combined}
\end{figure}

\begin{figure}
  \centering
  \begin{subfigure}[b]{0.45\textwidth}
    \includegraphics[width=\columnwidth]{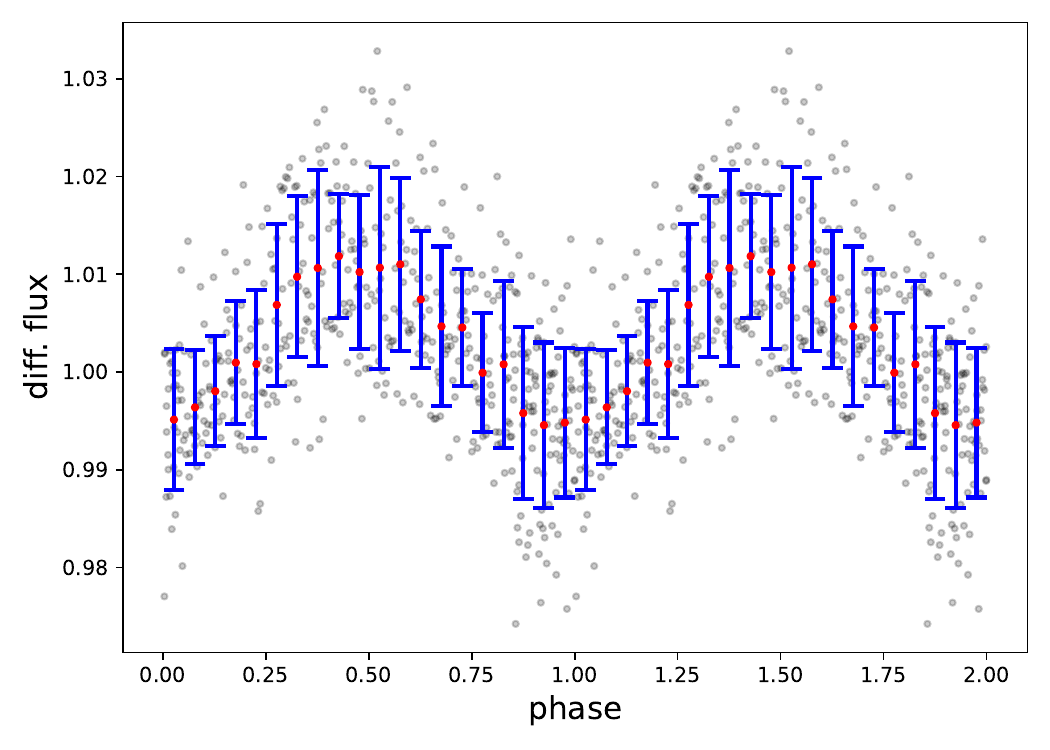}
  \end{subfigure}
  \hfill
  \begin{subfigure}[b]{0.45\textwidth}
    \includegraphics[width=\columnwidth]{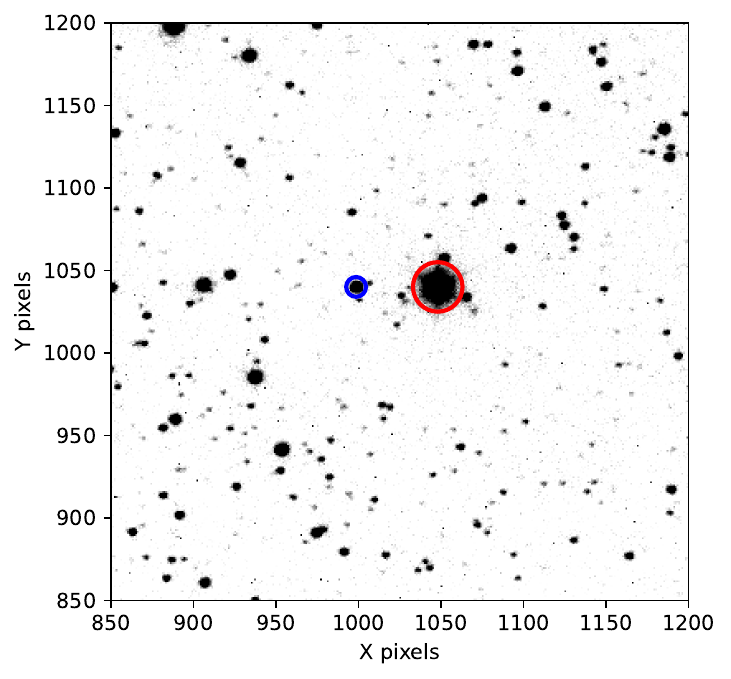}
  \end{subfigure}
    \caption{Top panel: The ground-based phased light curve of the contaminating star observed with TRAPPIST-South folded with the contaminating period. The red dots show the binned light curve, while the blue error bars are standard deviations in each bin. The grey dots in the background are the original phased light curve. Bottom: A cut-out of R-band CCD image in the field of HD\,100357 taken with TRAPPIST-South. The larger red circle to the right is HD\,100357 and the smaller blue circle to the left is Gaia DR3 5236626819707864192, which contaminates HD\,100357 in TESS pixels.}
  \label{fig:ground}
\end{figure}

    


\bsp	
\label{lastpage}
\end{document}